\documentclass[aps,prd,twocolumn,groupedaddress,amsmath,amssymb]{revtex4-1}
\usepackage{graphicx}  
\usepackage{dcolumn}   
\usepackage{bm}        
\usepackage{verbatim}   
\usepackage{subfigure}
\usepackage[utf8]{inputenc}
\usepackage{color}
\usepackage{braket}

\begin{document}

\title{Production and polarization of prompt $J/\psi$ in the improved color evaporation model using the $k_T$-factorization approach}
\author{Vincent Cheung}
\affiliation{
   Department of Physics,
   University of California, Davis,
   Davis, California 95616, USA
   }
\author{Ramona Vogt}
\affiliation{
   Nuclear and Chemical Sciences Division,
   Lawrence Livermore National Laboratory,
   Livermore, California 94551, USA
   }
\affiliation{
   Department of Physics,
   University of California, Davis,
   Davis, California 95616, USA
   }
\date{\today}
\begin{abstract}
We calculate the polarization of prompt $J/\psi$ production in the improved color evaporation model at leading order employing the $k_T$-factorization approach. In this paper, we present the polarization parameter $\lambda_\vartheta$ of prompt $J/\psi$ as a function of transverse momentum in $p+p$ and $p$ + A collisions to compare with data in the helicity, Collins-Soper and Gottfried-Jackson frames. We also present calculations of the charmonium production cross sections as a function of rapidity and transverse momentum. This is the first $p_T$-dependent calculation of charmonium polarization in the improved color evaporation model. We find agreement with both charmonium cross sections and polarization measurements.
\end{abstract}

\pacs{
14.40.Pq
}
\keywords{
Heavy Quarkonia}

\maketitle


\section{Introduction}
The production mechanism of quarkonium remains uncertain even more than 40 years after the discovery of the $J/\psi$. Nonrelativistic QCD (NRQCD) \cite{Caswell:1985ui}, the most widely employed model of quarkonium production encounters serious challenges in both the universality of the long distance matrix elements (LDMEs) and the prediction of quarkonium polarization \cite{Brambilla:2014jmp}. The production cross sections in NRQCD, based on an expansion in the strong coupling constant and the $Q\bar{Q}$ velocity \cite{Bodwin:1994jh}, is factorized into hard and soft contributions and divided into different color and spin states, including color octet contributions. The LDMEs, which weight the contributions from each color and spin state, are fit to the data above some minimum transverse momentum, $p_T$. These LDMEs, which are conjectured to be universal, fail to describe the yields and polarization simultaneously for $p_T$ cuts less than twice the mass of the quarkonium state \cite{Bodwin:2014gia,Faccioli:2014cqa}. They also depend on the collision system \cite{Ma:2010yw,Butenschoen:2010rq,Gong:2012ug,Zhang:2009ym}. Moreover, the polarization predicted by NRQCD is sensitive to the $p_T$ cut. The $\eta_c$ $p_T$ distributions calculated with LDMEs obtained from $J/\psi$ yields using heavy quark spin symmetry \cite{HQSS1,HQSS2,HQSS3}, generally overestimates the high $p_T$ LHCb $\eta_c$ results \cite{Butenschoen:2014dra}. NRQCD also consistently underestimates the $\Upsilon$($n$S) cross section ratio for 8 to 7~TeV as a function of $p_T$ \cite{Aaij:2015awa}.

On the other hand, the color evaporation model (CEM) \cite{Barger:1979js,Barger:1980mg,Gavai:1994in,Ma:2016exq}, which considers all $Q\bar{Q}$ ($Q$ = $c$, $b$) production regardless of the quark color, spin, and momentum, is able to predict both the total yields and the rapidity distributions with only a single normalization parameter per state \cite{NVF}. Both the CEM and NRQCD can describe production yields rather well but spin-related measurements like the polarization are strong tests of production models. However, polarization is not the only test of models. The CEM was also used recently to calculate transverse single spin asymmetries in $J/\psi$ production \cite{Godbole:2012bx,Godbole:2014tha}.

We have previously presented the first polarization results in the CEM \cite{Cheung:2017loo}, which only considered charmonium and bottomonium production in general, followed by polarization results of the prompt $J/\psi$ and $\Upsilon$(1S) \cite{Cheung:2017osx}. The later also took the feed-down production into account using the recently-developed improved CEM (ICEM) \cite{Ma:2016exq}. However, those results were at LO assuming collinear factorization and were thus $p_T$-independent. This paper serves as a continuation of the previous work by presenting a $p_T$-dependent leading order (LO) ICEM calculation of the polarization in prompt $J/\psi$ production using the $k_T$-factorization approach. This is a $p_T$-dependent result because the transverse momenta of the incoming gluons and their off-shell properties are not neglected in the $k_T$-factorization approach. Our calculation provides the first $p_T$-dependent ICEM polarization result and represents a step toward a full NLO ICEM polarization result. We will begin to address the $p_T$ dependence at NLO in a later publication.

In this paper, we present both the yields and polarizations of charmonium as a function of $p_T$ by formulating the ICEM in the $k_T$-factorization approach. In the high-energy limit, the contributions from $t$-channel gluon exchange can become dominant. The QCD evolution of the gluon distribution functions of the colliding partons is described by the BKFL evolution equation \cite{bkfl}. In this regime, the transverse momentum ($k_T$) of the incoming gluon can no longer be neglected. This phenomenological framework dealing with Reggeized $t$-channel gluons, is known as the $k_T$-factorization approach. We take the same effective Feynman rules for scattering processes involving incoming off-shell gluons \cite{Collins:1991ty} as in NRQCD \cite{Kniehl:2006sk}. Effectively, the momentum of the incoming Reggeon, $k^\mu$, with transverse momentum $k_T$ can be written in terms of the proton momentum $p^\mu$ and the fraction of longitudinal momentum $x$ carried by the gluon as 
\begin{eqnarray}
k^{\mu} = x p^\mu + k_{T}^\mu \;. \label{Reggeon_momentum}
\end{eqnarray}
The polarization 4-vector is
\begin{eqnarray}
\epsilon^\mu(k_{T}) = \frac{k_T^\mu}{k_T} \label{Reggeon_polarization} \;,
\end{eqnarray}
where $k_T^\mu = (0,\vec{k}_T,0)$.

In the traditional CEM, all quarkonium states are treated the same as $Q\bar{Q}$ below the $H\bar{H}$ threshold. The invariant mass of the heavy quark-antiquark pair is restricted to be less than twice the mass of the lowest mass meson ($H$) that can be formed with the heavy quark as a constituent. The distributions for all quarkonium family members are assumed to be identical.

In the ICEM the invariant mass of the intermediate heavy quark-antiquark pair is constrained to be larger than the mass of produced quarkonium state, $M_{\mathcal{Q}}$, instead of twice the quark mass, $2m_q$, the lower limit in the traditional CEM \cite{Barger:1979js,Cheung:2017loo}. Because the charmonium momentum and integration range depend on the mass of the state, the kinematic distributions of the charmonium states are no longer identical in the ICEM and, for example the $\psi^\prime$ to $J/\psi$ ratio depends on $p_T$. Using the $k_T$-factorization approach, in a $p+p$ collision, the ICEM production cross section for a directly-produced quarkonium state $\mathcal{Q}$ is
\begin{widetext}
\begin{eqnarray}
\sigma &=& F_\mathcal{Q} { \int_{ M_\mathcal{Q}^2}^{4m_H^2} d\hat{s} } \int \frac{dx_1}{x_1} \int \frac{d\phi_1}{2\pi} \int {dk_{1T}}^2 \Phi_1(x_1,k_{1T},\mu_{F1}^2) \int \frac{dx_2}{x_2} \int \frac{d\phi_2}{2\pi} \int{dk_{2T}}^2 \Phi_2(x_2,k_{2T},\mu_{F2}^2) \hat{\sigma}(R+R\rightarrow Q\bar{Q}) \nonumber \\
&\times& \delta(\hat{s} - x_1x_2 s +|\vec{k}_{1T}+\vec{k}_{2T}|^2) \;,
\label{cem_sigma}
\end{eqnarray}
\end{widetext}
where the square of the heavy quark pair invariant mass is $\hat{s}$ while the square of the center-of-mass energy in the $p+p$ collision is $s$. Here $\Phi(x,k_{T},\mu_F^2)$ is the unintegrated parton distribution function (uPDF) for a parton with momentum fraction $x$ and transverse momentum $k_T$ interacting with factorization scale $\mu_F$. The angles $\phi_{1,2}$ in Eq.~({\ref{cem_sigma}) are between the $k_{T1,2}$ of the partons and the $p_T$ of the final state quarkonium $\mathcal{Q}$. The parton-level cross section is $\sigma(R+R\rightarrow Q\bar{Q})$. Finally, $F_{\mathcal{Q}}$ is a universal factor for the directly-produced quarkonium state $\mathcal{Q}$, and is independent of the projectile, target, and energy. In this approach, the cross section is
\begin{widetext}
\begin{eqnarray}
\frac{d^{4}\sigma}{dp_{T} dy d\hat{s} d\phi} &=& \sigma \delta(\hat{s}-x_1 x_2 s + p_T^2) \delta \Big(y-\frac{1}{2}\log\frac{x_1}{x_2} \Big) \delta \Big(p_T^2-|\vec{k}_{1T}^2+\vec{k}_{2T}^2| \Big) \delta(\phi-(\phi_1-\phi_2)) \nonumber \\
&=& F_\mathcal{Q} \int \frac{2}{\pi} k_{2T} dk_{2T} \sum_{k_{1T}} \Bigg[\frac{\Phi_{1}(k_{1T},x_{10},\mu_{F1}^2)}{x_{10}} \frac{\Phi_{2}(k_{2T},x_{20},\mu_{F2}^2)}{x_{20}} k_{1T} p_T \frac{\hat{\sigma}(R+R\rightarrow Q\bar{Q})}{s \sqrt{k_{2T}^2(\cos^2\phi-1)+p_T^2}} \Bigg] \;
\label{int_rapidity}
\end{eqnarray}
\end{widetext}
where the sum $k_{1T}$ is over the roots of $k_{1T}^2+k_{2T}^2+2k_{1T}k_{2T}\cos\phi=p_T^2$, and $k_{1T,1}$, $k_{1T,2}$ are
\begin{eqnarray}
k_{1T,1} &=& -k_{2T}\cos \phi +\sqrt{k_{2T}^2(\cos^2\phi-1)+p_T^2} \label{k11} \\
k_{1T,2} &=& -k_{2T}\cos \phi -\sqrt{k_{2T}^2(\cos^2\phi-1)+p_T^2} \label{k12} \;.
\end{eqnarray}
The momentum fractions $x_{10}$ and $x_{20}$ are 
\begin{eqnarray}
x_{10} &=& \sqrt{\frac{\hat{s}+p_T^2}{s}}e^{+y} \;, \\
x_{20} &=& \sqrt{\frac{\hat{s}+p_T^2}{s}}e^{-y} \;.
\end{eqnarray}
Here, $\phi$ is the relative azimuthal angle between two incident Reggeons ($\phi=\phi_1-\phi_2$) and $p_T$ is the transverse momentum of the produced $Q\bar{Q}$.

The cross section may also be defined in terms of $x_F$ instead of rapidity as,
\begin{widetext}
\begin{eqnarray}
\frac{d^{4}\sigma}{dp_{T} dx_F d\hat{s} d\phi} &=& \sigma \delta(\hat{s}-x_1 x_2 s + p_T^2) \delta(x_F-(x_1-x_2)) \delta \Big(p_T^2-|\vec{k}_{1T}^2+\vec{k}_{2T}^2| \Big) \delta(\phi-(\phi_1-\phi_2)) \nonumber \\
&=& F_\mathcal{Q} \int \frac{2}{\pi} k_{2T} dk_{2T} \sum_{k_{1T}} \Bigg[\frac{\Phi_{1}(k_{1T},x_{10},\mu_{F1}^2)}{x_{10}} \frac{\Phi_{2}(k_{2T},x_{20},\mu_{F2}^2)}{x_{20}}k_{1T} p_T \nonumber \\
&\times& \frac{\hat{\sigma}(R+R\rightarrow Q\bar{Q})}{\sqrt{x_F^2 s^2+4(\hat{s}+p_T^2)} \sqrt{k_{2T}^2(\cos^2\phi-1)+p_T^2}} \Bigg] \;,
\label{int_xf}
\end{eqnarray}
\end{widetext}
where $x_{10}$ and $x_{20}$ are now
\begin{eqnarray}
x_{10} &=& \frac{1}{2}\Bigg(x_F+\sqrt{x_F^2+4\frac{\hat{s}+p_T^2}{s}}\Bigg) \\
x_{20} &=& \frac{1}{2}\Bigg(-x_F+\sqrt{x_F^2+4\frac{\hat{s}+p_T^2}{s}}\Bigg) \;.
\end{eqnarray}

Thus the transverse momentum distribution $d\sigma/dp_T$ in the ICEM is
\begin{eqnarray}
\label{cem_pt}
\frac{d\sigma}{dp_T} &=& \int dy d\hat{s} d\phi \frac{d^{4}\sigma}{dp_{T} dy d\hat{s} d\phi} \label{pt_rap_cut} \\
&=& \int dx_{F} d\hat{s} d\phi \frac{d^{4}\sigma}{dp_{T} dx_F d\hat{s} d\phi} \label{pt_xf_cut} \;.
\end{eqnarray}
The two expressions are equivalent when calculating the transverse momentum without any longitudinal kinematic cuts. Eq.~(\ref{pt_rap_cut}) is used to compare to collider data with defined rapidity cuts while Eq.~(\ref{pt_xf_cut}) is used to compare to fixed-target data with $x_F$ cuts. Similarly, the rapidity distribution $d\sigma/dy$ in the ICEM is
\begin{eqnarray}
\label{cem_y}
\frac{d\sigma}{dy} &=& \int dp_T d\hat{s} d\phi \frac{d^{4}\sigma}{dp_{T} dy d\hat{s} d\phi} \;. \label{y_pt_cut}
\end{eqnarray}

We take the renormalization and factorization scales to be $\mu_F=\mu_R=m_T$, where $m_T$ is the transverse mass of the $Q\bar{Q}$. We will study the effect of varying these scales on the $p_T$ distributions and the polarization.


\section{Polarization of directly produced $Q\bar{Q}$}
We define the polarization axis ($z$ axis) in the helicity frame where $z_{HX}$ is the flight direction of the quarkonium in the center of mass frame of the colliding beams, as shown in Fig.~\ref{polarization}. In this section we outline the kinematics required to compute the polarized scattering cross sections in the helicity frame as well as the procedure to relate them to the polarized scattering cross sections in the Gottfried-Jackson frame \cite{Gottfried:1964nx} and the Collins-Soper frame \cite{Collins:1977iv}.

In the lab frame, using Eqs.~(\ref{Reggeon_momentum}) and (\ref{Reggeon_polarization}) the momenta of the initial state Reggeons can be written as
\begin{eqnarray}
k_1^\mu &=& (x_1 s, k_{1T}\cos{\phi_1}, k_{1T}\sin{\phi_1},x_1 s) \\
k_2^\mu &=& (x_2 s, k_{2T}\cos{\phi_2}, k_{2T}\sin{\phi_2},-x_2 s) \;,
\end{eqnarray}
with polarization vectors
\begin{eqnarray}
\label{polarization_1}
\epsilon_1^\mu &=& \Big(0, \frac{\vec{k}_{1T}}{k_{1T}}, 0 \Big) = (0, \cos{\phi_1}, \sin{\phi_1}, 0 ) \\
\label{polarization_2} 
\epsilon_2^\mu &=& \Big(0, \frac{\vec{k}_{2T}}{k_{2T}}, 0 \Big) = (0, \cos{\phi_2}, \sin{\phi_2}, 0 ) \;.
\end{eqnarray}
We then boost the momenta along the beam direction to the frame where the total momentum of the Reggeons along the beam direction, $k_{1z}+k_{2z}$, is zero
\begin{eqnarray}
k_1^\mu &=& \Bigg(\frac{\sqrt{\hat{s}+p_T^2}}{2}, \vec{k}_{1T}, \frac{\sqrt{\hat{s}+p_T^2}}{2} \Bigg) \;, \\
k_2^\mu &=& \Bigg(\frac{\sqrt{\hat{s}+p_T^2}}{2}, \vec{k}_{2T}, -\frac{\sqrt{\hat{s}+p_T^2}}{2} \Bigg) \;,
\end{eqnarray}
where $\hat{s}=x_1 x_2 s - |\vec{k}_{1T}+\vec{k}_{2T}|^2$ and $p_T^2 = |\vec{k}_{1T}+\vec{k}_{2T}|^2$. The polarization vectors are unchanged. We then apply a rotation such that the three momentum of the sum $k_1^\mu+k_2^\mu$ is aligned with a new $z$-axis 
\begin{eqnarray}
k_1^\mu+k_2^\mu &=& \Big(\sqrt{p_T^2+\hat{s}},\vec{0},p_T\Big) \;.
\end{eqnarray}
We then boost to the quarkonium rest frame where
\begin{eqnarray}
k_1^\mu+k_2^\mu &=& \Big(\sqrt{\hat{s}},\vec{0},0\Big) \;.
\end{eqnarray}
In this frame (helicity frame), the momenta of the initial state Reggeons are 
\begin{eqnarray}
k_1^\mu &=& \Big(\frac{-\psi+\hat{s}}{2\sqrt{\hat{s}}},\frac{\sqrt{\hat{s}}\lambda}{2}, \frac{k_{1T} k_{2T} \sin \phi}{p_T},\frac{\psi \lambda}{2p_T} \Big) \;, \\
k_2^\mu &=& \Big(\frac{\psi+\hat{s}}{2\sqrt{\hat{s}}},-\frac{\sqrt{\hat{s}}\lambda}{2}, -\frac{k_{1T} k_{2T}| \sin \phi}{p_T},-\frac{\psi \lambda}{2p_T} \Big) \;,
\end{eqnarray}
where $\psi=|\vec{k}_{1T}|^2-|\vec{k}_{2T}|^2$, $\phi=\phi_1-\phi_2$, and $\lambda=\sqrt{1+p_T^2/\hat{s}}$. The polarization vectors are now
\begin{eqnarray}
\epsilon_1^\mu &=& \Big(-\frac{k_{1T}+k_{2T}\cos \phi}{\sqrt{\hat{s}}},0,\frac{k_{2T}\sin \phi}{p_T}, \\
&&\frac{\lambda}{p_T} (k_{1T}+k_{2T}\cos \phi) \Big) \;, \\
\epsilon_2^\mu &=& \Big(-\frac{k_{2T}+k_{1T}\cos \phi}{\sqrt{\hat{s}}},0,-\frac{k_{1T}\sin \phi}{p_T}, \\
&&\frac{\lambda}{p_T} (k_{2T}+k_{1T}\cos \phi) \Big) \;.
\end{eqnarray}

\begin{figure}
\centering
\includegraphics[width=\columnwidth]{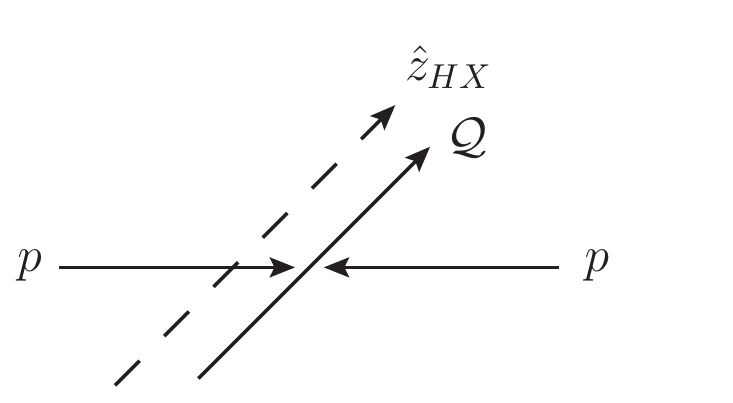}
\caption{\label{polarization} The orientation of polarization axis ($z$ axis) in the helicity frame is indicated by the dashed arrow. The proton arrows indicate the incoming beam directions. The polarization axis is defined to be the direction of the produced ($\mathcal{Q}$) travels in the center-of-mass frame of the colliding beams. If the quarks in the $Q\bar{Q}$ pair with total angular momentum $J=1$, they can either have the same angular momentum along the $z$-axis, $J_z$, or opposite resulting in $J_z=0$ (longitudinal) or $J_z=1$ (transverse), respectively.}
\end{figure}

The scattering amplitude of the process $R+R\rightarrow Q\bar{Q}$ is related to that of $g+g\rightarrow Q\bar{Q}$ by \cite{Collins:1991ty,Kniehl:2006sk}
\begin{eqnarray}
\mathcal{A}(R+R\rightarrow Q + \bar{Q}) &=& \epsilon^\mu(k_1) \epsilon^\nu(k_2) \\
&\times&\mathcal{A}_{\mu \nu}(g+g\rightarrow Q + \bar{Q})  \;,
\end{eqnarray}
where $\epsilon^\mu(k)$ is defined in Eq.~(\ref{Reggeon_polarization}). Evaluating $\mathcal{A}_{\mu \nu}(g+g\rightarrow Q + \bar{Q})$ using the conventional Feynman rules of QCD, there are three $gg \rightarrow Q\bar{Q}$ Feynman diagrams to consider at $\mathcal{O}(\alpha_s^2)$. Each diagram includes a color factor $C$ and a scattering amplitude $\mathcal{A}$. The generic matrix element for the gluon fusion process can be written as \cite{Cvitanovic:1976am}
\begin{eqnarray}
\mathcal{M}_{gg} &=& C_{gg,\hat{s}} \mathcal{A}_{gg,\hat{s}} + C_{gg,\hat{t}} \mathcal{A}_{gg,\hat{t}} + C_{gg,\hat{u}} \mathcal{A}_{gg,\hat{u}} \;.
\label{matrixelement}
\end{eqnarray}
In terms of the Dirac spinors $u$ and $v$, the individual amplitudes are
\begin{eqnarray}
\mathcal{A}_{gg,\hat{s}} &=& - \frac{g_s^2}{\hat{s}} \Big\{ -(2k^\prime+k) \cdot \epsilon(k) [\bar{u}(p^\prime) \epsilon\!\!\!/(k^\prime) v(p)] \nonumber \\
&+&(2k+k^\prime)\cdot \epsilon(k^\prime) [\bar{u}(p^\prime) \epsilon\!\!\!/(k) v(p)] \nonumber \\
&+& \epsilon(k) \cdot\epsilon(k^\prime)[\bar{u}(p^\prime) (k\!\!\!/^\prime - k\!\!\!/) v(p)]\Big\} \;, \\
\mathcal{A}_{gg,\hat{t}} &=& -\frac{g_s^2}{\hat{t}-m_c^2} \bar{u}(p^\prime) \epsilon\!\!\!/(k^\prime) (k\!\!\!/ -p\!\!\!/ +m_c) \epsilon\!\!\!/(k) v(p) \;, \\
\mathcal{A}_{gg,\hat{u}} &=& -\frac{g_s^2}{\hat{u}-m_c^2} \bar{u}(p^\prime) \epsilon\!\!\!/(k) (k\!\!\!/^\prime -p\!\!\!/ +m_c) \epsilon\!\!\!/(k^\prime) v(p) \;.
\end{eqnarray}
Here $g_s$ is the gauge coupling, $m_c$ is the charm quark mass, $\epsilon$ represents the gluon polarization vectors, $\gamma^\mu$ are the gamma matrices, $k^\prime$ ($k$) is the momentum of the initial state light quark (antiquark) or gluon, and $p^\prime$ ($p$) is the momentum of final state heavy quark (antiquark). 

In the process of evaluating the scattering amplitudes, we take advantage of the fact, that at $\mathcal{O}(\alpha_s^2)$, the final state $Q\bar{Q}$ is produced with no dependence on the azimuthal angle $\phi^\prime$ (and thus $L_{z^\prime}=0$) in a rotated frame (primed frame) where the $z^\prime$ axis is defined as the direction of one of the incoming Reggeons. Since the Reggeons are head to head in this frame, the scattering amplitudes are independent of the azimuthal angle $\phi^\prime$. We first rotate the initial state momenta $\vec{p}$ from the helicity frame to the primed frame by an Euler rotation:
\begin{eqnarray}
\label{forward_rotation}
\vec{p^\prime} &=& \mathcal{R}(0,\beta,\gamma) \vec{p}.
\end{eqnarray}

The scattering amplitudes in the primed frame for $S=1$, sorted $S_{z^\prime}$, are
\begin{eqnarray}
\mathcal{A}_{gg,\hat{s},S=1,S_{z^\prime}=0} &=& \frac{1}{\sqrt{2}}[(\mathcal{A}_{s1})+(\mathcal{A}_{s4})] \;, \label{s_triplet} \\
\mathcal{A}_{gg,\hat{s},S=1,S_{z^\prime}=\pm1} &=& \mathcal{A}_{s2,3} \;, \\
\mathcal{A}_{gg,\hat{t},S=1,S_{z^\prime}=0} &=& \frac{1}{\sqrt{2}} [(\mathcal{A}_{t1})+(\mathcal{A}_{t4})]  \;, \label{t_triplet}\\
\mathcal{A}_{gg,\hat{t},S=1,S_{^\prime}=\pm1} &=& \mathcal{A}_{t2,3} \;, \\
\mathcal{A}_{gg,\hat{u},S=1,S_{z^\prime}=0} &=& \frac{1}{\sqrt{2}} [(\mathcal{A}_{u1})+(\mathcal{A}_{u4})] \;, \label{u_triplet} \\
\mathcal{A}_{gg,\hat{u},S=1,S_{^\prime}=\pm1} &=& \mathcal{A}_{u2,3} \;.
\end{eqnarray}

The $\hat{s}$-channel amplitudes are:
\begin{widetext}
\begin{eqnarray}
\mathcal{A}_{s1} \frac{\hat{s}}{g_s^2} &=& \frac{m_c}{\sqrt{\hat{s}}}\Big[ [(-\psi+3\hat{s}) \epsilon_1^3 \epsilon_2^0 + (-\psi-3\hat{s}) \epsilon_1^0 \epsilon_2^3 + (-2 \epsilon_1^0 \epsilon_2^0+2\epsilon_1^1 \epsilon_2^1+2\epsilon_1^2 \epsilon_2^2) \sqrt{(k_{1T} - k_{2T})^2+\hat{s})(k_{1T} + k_{2T})^2+\hat{s})}]\cos \theta^\prime \nonumber \\
&-& [(\psi-3\hat{s}) \epsilon_1^1 \epsilon_2^0 + (\psi+3\hat{s}) \epsilon_1^0 \epsilon_2^1 + (\epsilon_1^3 \epsilon_2^1 + \epsilon_1^1 \epsilon_2^3) \sqrt{(k_{1T} - k_{2T})^2+\hat{s})(k_{1T} + k_{2T})^2+\hat{s})}]\sin \theta^\prime \Big]\;,
\end{eqnarray}
\end{widetext}
\begin{widetext}
\begin{eqnarray}
\mathcal{A}_{s2} \frac{\hat{s}}{g_s^2} &=& \frac{i}{2}[(\psi-3\hat{s})\epsilon_1^2 \epsilon_2^0 + (\psi+3\hat{s}) \epsilon_1^0 \epsilon_2^2 + (\epsilon_1^3 \epsilon_2^2 + \epsilon_1^2 \epsilon_2^3) \sqrt{(k_{1T} - k_{2T})^2+\hat{s})(k_{1T} + k_{2T})^2+\hat{s})}] \nonumber \\
&-& \frac{1}{2}[(\psi-3\hat{s})\epsilon_1^1 \epsilon_2^0 + (\psi+3\hat{s}) \epsilon_1^0 \epsilon_2^1 + (\epsilon_1^3 \epsilon_2^1 + \epsilon_1^1 \epsilon_2^3) \sqrt{(k_{1T} - k_{2T})^2+\hat{s})(k_{1T} + k_{2T})^2+\hat{s})}]\cos \theta^\prime \nonumber \\
&+& \frac{1}{2}[(\psi-3\hat{s})\epsilon_1^3 \epsilon_2^0 + (\psi+3\hat{s}) \epsilon_1^0 \epsilon_2^3 + (2\epsilon_1^0 \epsilon_2^0 -2 \epsilon_1^1 \epsilon_2^1 -2 \epsilon_1^2 \epsilon_2^2) \sqrt{(k_{1T} - k_{2T})^2+\hat{s})(k_{1T} + k_{2T})^2+\hat{s})}]\sin \theta^\prime \;, \nonumber \\
\end{eqnarray}
\end{widetext}
\begin{widetext}
\begin{eqnarray}
\mathcal{A}_{s3} \frac{\hat{s}}{g_s^2} &=& -\frac{i}{2}[(\psi-3\hat{s})\epsilon_1^2 \epsilon_2^0 + (\psi+3\hat{s}) \epsilon_1^0 \epsilon_2^2 + (\epsilon_1^3 \epsilon_2^2 + \epsilon_1^2 \epsilon_2^3) \sqrt{(k_{1T} - k_{2T})^2+\hat{s})(k_{1T} + k_{2T})^2+\hat{s})}] \nonumber \\
&-&\frac{1}{2}[(\psi-3\hat{s})\epsilon_1^1 \epsilon_2^0 + (\psi+3\hat{s}) \epsilon_1^0 \epsilon_2^1 + (\epsilon_1^3 \epsilon_2^1 + \epsilon_1^1 \epsilon_2^3) \sqrt{(k_{1T} - k_{2T})^2+\hat{s})(k_{1T} + k_{2T})^2+\hat{s})}]\cos \theta^\prime \nonumber \\
&+&\frac{1}{2} [(\psi-3\hat{s})\epsilon_1^3 \epsilon_2^0 + (\psi+3\hat{s}) \epsilon_1^0 \epsilon_2^3 + (2\epsilon_1^0 \epsilon_2^0 - 2\epsilon_1^1 \epsilon_2^1 - 2\epsilon_1^2 \epsilon_2^2) \sqrt{(k_{1T} - k_{2T})^2+\hat{s})(k_{1T} + k_{2T})^2+\hat{s})}]\sin \theta^\prime \;, \nonumber \\
\end{eqnarray}
\end{widetext}
\begin{widetext}
\begin{eqnarray}
\mathcal{A}_{s4} \frac{\hat{s}}{g_s^2} &=& \frac{m_c}{\sqrt{\hat{s}}}\Big[ [(\psi-3\hat{s}) \epsilon_1^3 \epsilon_2^0 + (\psi+3\hat{s}) \epsilon_1^0 \epsilon_2^3 + (2\epsilon_1^0 \epsilon_2^0 - 2\epsilon_1^1 \epsilon_2^1 - 2\epsilon_1^2 \epsilon_2^2) \sqrt{(k_{1T} - k_{2T})^2+\hat{s})(k_{1T} + k_{2T})^2+\hat{s})}]\cos \theta^\prime \nonumber \\
&+& [(\psi-3\hat{s}) \epsilon_1^1 \epsilon_2^0 + (\psi+3\hat{s}) \epsilon_1^0 \epsilon_2^1 + (\epsilon_1^3 \epsilon_2^1 + \epsilon_1^1 \epsilon_2^3) \sqrt{(k_{1T} - k_{2T})^2+\hat{s})(k_{1T} + k_{2T})^2+\hat{s})}]\sin \theta^\prime \Big] \;.
\end{eqnarray}
\end{widetext}

The $\hat{t}$-channel amplitudes are:
\begin{widetext}
\begin{eqnarray}
\mathcal{A}_{t1} \frac{\hat{t}-m_c^2}{g_s^2} &=& -2 \epsilon_1^1 \epsilon_2^1 m_c \sqrt{\hat{s}} \chi + i \frac{(\epsilon_1^2 \epsilon_2^1 - \epsilon_1^1 \epsilon_2^2)m_c\sqrt{(k_{1T} - k_{2T})^2+\hat{s})(k_{1T} + k_{2T})^2+\hat{s})}}{\sqrt{\hat{s}}} \nonumber \\ 
&-& \frac{m_c}{\sqrt{\hat{s}}}\Big[ (\psi-\hat{s}) \epsilon_1^3 \epsilon_2^0 + (\psi+\hat{s}) \epsilon_1^0 \epsilon_2^3 + (\epsilon_1^0 \epsilon_2^0 - \epsilon_1^1 \epsilon_2^1 - \epsilon_1^2 \epsilon_2^2 + \epsilon_1^3 \epsilon_2^3) \sqrt{(k_{1T} - k_{2T})^2+\hat{s})(k_{1T} + k_{2T})^2+\hat{s})} \Big] \cos\theta^\prime \nonumber \\
&+& 2 m_c \sqrt{\hat{s}} \chi (\epsilon_1^1 \epsilon_2^1 - \epsilon_1^3 \epsilon_2^3 ) \cos^2\theta^\prime \nonumber \\
&-&\frac{m_c}{\sqrt{\hat{s}}}\Big[ (\psi-\hat{s}) \epsilon_1^1 \epsilon_2^0 + (\psi+\hat{s}) \epsilon_1^0 \epsilon_2^1 + (\epsilon_1^3 \epsilon_2^1 + \epsilon_1^1 \epsilon_2^3) \sqrt{(k_{1T} - k_{2T})^2+\hat{s})(k_{1T} + k_{2T})^2+\hat{s})} \Big] \sin\theta^\prime \nonumber \\
&-& 2 m_c \sqrt{\hat{s}} \chi (\epsilon_1^3 \epsilon_2^1 + \epsilon_1^1 \epsilon_2^3) \sin\theta^\prime \cos\theta^\prime \;,
\end{eqnarray}
\end{widetext}
\begin{widetext}
\begin{eqnarray}
\mathcal{A}_{t2} \frac{\hat{t}-m_c^2}{g_s^2} &=& -\frac{1}{2} \Big[ (\psi-\hat{s}) \epsilon_1^3 \epsilon_2^1 - (\psi+\hat{s}) \epsilon_1^1 \epsilon_2^3 - (\epsilon_1^1 \epsilon_2^0 - \epsilon_1^0 \epsilon_2^1 ) \sqrt{(k_{1T} - k_{2T})^2+\hat{s})(k_{1T} + k_{2T})^2+\hat{s})} \Big] \chi \nonumber \\
&+& i\frac{1}{2}\Big[ (\psi-\hat{s}) \epsilon_1^2 \epsilon_2^0 + (\psi+\hat{s}) \epsilon_1^0 \epsilon_2^2 + (\epsilon_1^3 \epsilon_2^2 + \epsilon_1^2 \epsilon_2^3) \sqrt{(k_{1T} - k_{2T})^2+\hat{s})(k_{1T} + k_{2T})^2+\hat{s})} \Big] \nonumber \\
&-& \frac{1}{2} \Big[ (\psi-\hat{s}) \epsilon_1^1 \epsilon_2^0 + (\psi+\hat{s}) \epsilon_1^0 \epsilon_2^1 + (\epsilon_1^3 \epsilon_2^1 + \epsilon_1^1 \epsilon_2^3) \sqrt{(k_{1T} - k_{2T})^2+\hat{s})(k_{1T} + k_{2T})^2+\hat{s})} \Big]\cos\theta^\prime \nonumber \\
&-& i \frac{1}{2} \Big[ (\psi-\hat{s}) \epsilon_1^2 \epsilon_2^3 - (\psi+\hat{s}) \epsilon_1^3 \epsilon_2^2 + (\epsilon_1^2 \epsilon_2^0 - \epsilon_1^0 \epsilon_2^2) \sqrt{(k_{1T} - k_{2T})^2+\hat{s})(k_{1T} + k_{2T})^2+\hat{s})} \Big] \chi \cos\theta^\prime \nonumber \\
&-& (\epsilon_1^3 \epsilon_2^1 + \epsilon_1^1 \epsilon_2^3)\hat{s} \chi \cos^2\theta^\prime \nonumber\\
&+& \frac{1}{2} \Big[ (\psi-\hat{s}) \epsilon_1^3 \epsilon_2^0 + (\psi+\hat{s}) \epsilon_1^0 \epsilon_2^3 + (\epsilon_1^0 \epsilon_2^0 - \epsilon_1^1 \epsilon_2^1 - \epsilon_1^2 \epsilon_2^2 + \epsilon_1^3 \epsilon_2^3) \sqrt{(k_{1T} - k_{2T})^2+\hat{s})(k_{1T} + k_{2T})^2+\hat{s})} \Big]\sin\theta^\prime \nonumber \\
&-& i\frac{1}{2} \Big[ (\psi-\hat{s})\epsilon_1^2 \epsilon_2^1 - (\psi+\hat{s})\epsilon_1^1\epsilon_2^2 \Big]\chi \sin \theta^\prime - (\epsilon_1^1\epsilon_2^1-\epsilon_1^3\epsilon_2^3)\hat{s}\chi \sin\theta^\prime \cos\theta^\prime \;,
\end{eqnarray}
\end{widetext}
\begin{widetext}
\begin{eqnarray}
\mathcal{A}_{t3} \frac{\hat{t}-m_c^2}{g_s^2} &=& -\frac{1}{2} \Big[ (\psi-\hat{s}) \epsilon_1^3 \epsilon_2^1 - (\psi+\hat{s}) \epsilon_1^1 \epsilon_2^3 - (\epsilon_1^1 \epsilon_2^0 - \epsilon_1^0 \epsilon_2^1 ) \sqrt{(k_{1T} - k_{2T})^2+\hat{s})(k_{1T} + k_{2T})^2+\hat{s})} \Big] \chi \nonumber \\
&-& i\frac{1}{2}\Big[ (\psi-\hat{s}) \epsilon_1^2 \epsilon_2^0 + (\psi+\hat{s}) \epsilon_1^0 \epsilon_2^2 + (\epsilon_1^3 \epsilon_2^2 + \epsilon_1^2 \epsilon_2^3) \sqrt{(k_{1T} - k_{2T})^2+\hat{s})(k_{1T} + k_{2T})^2+\hat{s})} \Big] \nonumber \\
&-& \frac{1}{2} \Big[ (\psi-\hat{s}) \epsilon_1^1 \epsilon_2^0 + (\psi+\hat{s}) \epsilon_1^0 \epsilon_2^1 + (\epsilon_1^3 \epsilon_2^1 + \epsilon_1^1 \epsilon_2^3) \sqrt{(k_{1T} - k_{2T})^2+\hat{s})(k_{1T} + k_{2T})^2+\hat{s})} \Big]\cos\theta^\prime \nonumber \\
&+& i \frac{1}{2} \Big[ (\psi-\hat{s}) \epsilon_1^2 \epsilon_2^3 - (\psi+\hat{s}) \epsilon_1^3 \epsilon_2^2 + (\epsilon_1^2 \epsilon_2^0 - \epsilon_1^0 \epsilon_2^2) \sqrt{(k_{1T} - k_{2T})^2+\hat{s})(k_{1T} + k_{2T})^2+\hat{s})} \Big] \chi \cos\theta^\prime \nonumber \\
&-& (\epsilon_1^3 \epsilon_2^1 + \epsilon_1^1 \epsilon_2^3)\hat{s} \chi \cos^2\theta^\prime \nonumber\\
&+& \frac{1}{2} \Big[ (\psi-\hat{s}) \epsilon_1^3 \epsilon_2^0 + (\psi+\hat{s}) \epsilon_1^0 \epsilon_2^3 + (\epsilon_1^0 \epsilon_2^0 - \epsilon_1^1 \epsilon_2^1 - \epsilon_1^2 \epsilon_2^2 + \epsilon_1^3 \epsilon_2^3) \sqrt{(k_{1T} - k_{2T})^2+\hat{s})(k_{1T} + k_{2T})^2+\hat{s})} \Big]\sin\theta^\prime \nonumber \\
&+& i\frac{1}{2} \Big[ (\psi-\hat{s})\epsilon_1^2 \epsilon_2^1 - (\psi+\hat{s})\epsilon_1^1\epsilon_2^2 \Big]\chi \sin \theta^\prime - (\epsilon_1^1\epsilon_2^1-\epsilon_1^3\epsilon_2^3)\hat{s}\chi \sin\theta^\prime \cos\theta^\prime \;,
\end{eqnarray}
\end{widetext}
\begin{widetext}
\begin{eqnarray}
\mathcal{A}_{t4} \frac{\hat{t}-m_c^2}{g_s^2} &=& 2 \epsilon_1^1 \epsilon_2^1 m_c \sqrt{\hat{s}} \chi + i \frac{(\epsilon_1^2 \epsilon_2^1 - \epsilon_1^1 \epsilon_2^2)m_c\sqrt{(k_{1T} - k_{2T})^2+\hat{s})(k_{1T} + k_{2T})^2+\hat{s})}}{\sqrt{\hat{s}}} \nonumber \\ 
&+& \frac{m_c}{\sqrt{\hat{s}}}\Big[ (\psi-\hat{s}) \epsilon_1^3 \epsilon_2^0 + (\psi+\hat{s}) \epsilon_1^0 \epsilon_2^3 + (\epsilon_1^0 \epsilon_2^0 - \epsilon_1^1 \epsilon_2^1 - \epsilon_1^2 \epsilon_2^2 + \epsilon_1^3 \epsilon_2^3) \sqrt{(k_{1T} - k_{2T})^2+\hat{s})(k_{1T} + k_{2T})^2+\hat{s})} \Big] \cos\theta^\prime \nonumber \\
&-& 2 m_c \sqrt{\hat{s}} \chi (\epsilon_1^1 \epsilon_2^1 - \epsilon_1^3 \epsilon_2^3 ) \cos^2\theta^\prime \nonumber \\
&+&\frac{m_c}{\sqrt{\hat{s}}}\Big[ (\psi-\hat{s}) \epsilon_1^1 \epsilon_2^0 + (\psi+\hat{s}) \epsilon_1^0 \epsilon_2^1 + (\epsilon_1^3 \epsilon_2^1 + \epsilon_1^1 \epsilon_2^3) \sqrt{(k_{1T} - k_{2T})^2+\hat{s})(k_{1T} + k_{2T})^2+\hat{s})} \Big] \sin\theta^\prime \nonumber \\
&+& 2 m_c \sqrt{\hat{s}} \chi (\epsilon_1^3 \epsilon_2^1 + \epsilon_1^1 \epsilon_2^3) \sin\theta^\prime \cos\theta^\prime \;.
\end{eqnarray}
\end{widetext}

Finally, the $\hat{u}$-channel amplitudes are
\begin{widetext}
\begin{eqnarray}
\mathcal{A}_{u1} \frac{\hat{u}-m_c^2}{g_s^2} &=& -2 \epsilon_1^1 \epsilon_2^1 m_c \sqrt{\hat{s}} \chi + i \frac{(\epsilon_1^2 \epsilon_2^1 - \epsilon_1^1 \epsilon_2^2)M\sqrt{(k_{1T} - k_{2T})^2+\hat{s})(k_{1T} + k_{2T})^2+\hat{s})}}{\sqrt{\hat{s}}} \nonumber \\ 
&+& \frac{m_c}{\sqrt{\hat{s}}}\Big[ (\psi-\hat{s}) \epsilon_1^3 \epsilon_2^0 + (\psi+\hat{s}) \epsilon_1^0 \epsilon_2^3 + (\epsilon_1^0 \epsilon_2^0 - \epsilon_1^1 \epsilon_2^1 - \epsilon_1^2 \epsilon_2^2 + \epsilon_1^3 \epsilon_2^3) \sqrt{(k_{1T} - k_{2T})^2+\hat{s})(k_{1T} + k_{2T})^2+\hat{s})} \Big] \cos\theta^\prime \nonumber \\
&+& 2 m_c \sqrt{\hat{s}} \chi (\epsilon_1^1 \epsilon_2^1 - \epsilon_1^3 \epsilon_2^3 ) \cos^2\theta^\prime \nonumber \\
&+&\frac{m_c}{\sqrt{\hat{s}}}\Big[ (\psi-\hat{s}) \epsilon_1^1 \epsilon_2^0 + (\psi+\hat{s}) \epsilon_1^0 \epsilon_2^1 + (\epsilon_1^3 \epsilon_2^1 + \epsilon_1^1 \epsilon_2^3) \sqrt{(k_{1T} - k_{2T})^2+\hat{s})(k_{1T} + k_{2T})^2+\hat{s})} \Big] \sin\theta^\prime \nonumber \\
&-& 2 m_c \sqrt{\hat{s}} \chi (\epsilon_1^3 \epsilon_2^1 + \epsilon_1^1 \epsilon_2^3) \sin\theta^\prime \cos\theta^\prime \;,
\end{eqnarray}
\end{widetext}
\begin{widetext}
\begin{eqnarray}
\mathcal{A}_{u2} \frac{\hat{u}-m_c^2}{g_s^2} &=& -\frac{1}{2} \Big[ (\psi-\hat{s}) \epsilon_1^3 \epsilon_2^1 - (\psi+\hat{s}) \epsilon_1^1 \epsilon_2^3 - (\epsilon_1^1 \epsilon_2^0 - \epsilon_1^0 \epsilon_2^1 ) \sqrt{(k_{1T} - k_{2T})^2+\hat{s})(k_{1T} + k_{2T})^2+\hat{s})} \Big] \chi \nonumber \\
&-& i\frac{1}{2}\Big[ (\psi-\hat{s}) \epsilon_1^2 \epsilon_2^0 + (\psi+\hat{s}) \epsilon_1^0 \epsilon_2^2 + (\epsilon_1^3 \epsilon_2^2 + \epsilon_1^2 \epsilon_2^3) \sqrt{(k_{1T} - k_{2T})^2+\hat{s})(k_{1T} + k_{2T})^2+\hat{s})} \Big] \nonumber \\
&+& \frac{1}{2} \Big[ (\psi-\hat{s}) \epsilon_1^1 \epsilon_2^0 + (\psi+\hat{s}) \epsilon_1^0 \epsilon_2^1 + (\epsilon_1^3 \epsilon_2^1 + \epsilon_1^1 \epsilon_2^3) \sqrt{(k_{1T} - k_{2T})^2+\hat{s})(k_{1T} + k_{2T})^2+\hat{s})} \Big]\cos\theta^\prime \nonumber \\
&-& i \frac{1}{2} \Big[ (\psi-\hat{s}) \epsilon_1^2 \epsilon_2^3 - (\psi+\hat{s}) \epsilon_1^3 \epsilon_2^2 + (\epsilon_1^2 \epsilon_2^0 - \epsilon_1^0 \epsilon_2^2) \sqrt{(k_{1T} - k_{2T})^2+\hat{s})(k_{1T} + k_{2T})^2+\hat{s})} \Big] \chi \cos\theta^\prime \nonumber \\
&-& (\epsilon_1^3 \epsilon_2^1 + \epsilon_1^1 \epsilon_2^3)\hat{s} \chi \cos^2\theta^\prime \nonumber\\
&-& \frac{1}{2} \Big[ (\psi-\hat{s}) \epsilon_1^3 \epsilon_2^0 + (\psi+\hat{s}) \epsilon_1^0 \epsilon_2^3 + (\epsilon_1^0 \epsilon_2^0 - \epsilon_1^1 \epsilon_2^1 - \epsilon_1^2 \epsilon_2^2 + \epsilon_1^3 \epsilon_2^3) \sqrt{(k_{1T} - k_{2T})^2+\hat{s})(k_{1T} + k_{2T})^2+\hat{s})} \Big]\sin\theta^\prime \nonumber \\
&-& i\frac{1}{2} \Big[ (\psi-\hat{s})\epsilon_1^2 \epsilon_2^1 - (\psi+\hat{s})\epsilon_1^1\epsilon_2^2 \Big]\chi \sin \theta^\prime - (\epsilon_1^1\epsilon_2^1-\epsilon_1^3\epsilon_2^3)\hat{s}\chi \sin\theta^\prime \cos\theta^\prime \;,
\end{eqnarray}
\end{widetext}
\begin{widetext}
\begin{eqnarray}
\mathcal{A}_{u3} \frac{\hat{u}-m_c^2}{g_s^2} &=& -\frac{1}{2} \Big[ (\psi-\hat{s}) \epsilon_1^3 \epsilon_2^1 - (\psi+\hat{s}) \epsilon_1^1 \epsilon_2^3 - (\epsilon_1^1 \epsilon_2^0 - \epsilon_1^0 \epsilon_2^1 ) \sqrt{(k_{1T} - k_{2T})^2+\hat{s})(k_{1T} + k_{2T})^2+\hat{s})} \Big] \chi \nonumber \\
&+& i\frac{1}{2}\Big[ (\psi-\hat{s}) \epsilon_1^2 \epsilon_2^0 + (\psi+\hat{s}) \epsilon_1^0 \epsilon_2^2 + (\epsilon_1^3 \epsilon_2^2 + \epsilon_1^2 \epsilon_2^3) \sqrt{(k_{1T} - k_{2T})^2+\hat{s})(k_{1T} + k_{2T})^2+\hat{s})} \Big] \nonumber \\
&+& \frac{1}{2} \Big[ (\psi-\hat{s}) \epsilon_1^1 \epsilon_2^0 + (\psi+\hat{s}) \epsilon_1^0 \epsilon_2^1 + (\epsilon_1^3 \epsilon_2^1 + \epsilon_1^1 \epsilon_2^3) \sqrt{(k_{1T} - k_{2T})^2+\hat{s})(k_{1T} + k_{2T})^2+\hat{s})} \Big]\cos\theta^\prime \nonumber \\
&+& i \frac{1}{2} \Big[ (\psi-\hat{s}) \epsilon_1^2 \epsilon_2^3 - (\psi+\hat{s}) \epsilon_1^3 \epsilon_2^2 + (\epsilon_1^2 \epsilon_2^0 - \epsilon_1^0 \epsilon_2^2) \sqrt{(k_{1T} - k_{2T})^2+\hat{s})(k_{1T} + k_{2T})^2+\hat{s})} \Big] \chi \cos\theta^\prime \nonumber \\
&-& (\epsilon_1^3 \epsilon_2^1 + \epsilon_1^1 \epsilon_2^3)\hat{s} \chi \cos^2\theta^\prime \nonumber\\
&-& \frac{1}{2} \Big[ (\psi-\hat{s}) \epsilon_1^3 \epsilon_2^0 + (\psi+\hat{s}) \epsilon_1^0 \epsilon_2^3 + (\epsilon_1^0 \epsilon_2^0 - \epsilon_1^1 \epsilon_2^1 - \epsilon_1^2 \epsilon_2^2 + \epsilon_1^3 \epsilon_2^3) \sqrt{(k_{1T} - k_{2T})^2+\hat{s})(k_{1T} + k_{2T})^2+\hat{s})} \Big]\sin\theta^\prime \nonumber \\
&+& i\frac{1}{2} \Big[ (\psi-\hat{s})\epsilon_1^2 \epsilon_2^1 - (\psi+\hat{s})\epsilon_1^1\epsilon_2^2 \Big]\chi \sin \theta^\prime - (\epsilon_1^1\epsilon_2^1-\epsilon_1^3\epsilon_2^3)\hat{s}\chi \sin\theta^\prime \cos\theta^\prime \;,
\end{eqnarray}
\end{widetext}
\begin{widetext}
\begin{eqnarray}
\mathcal{A}_{u4} \frac{\hat{u}-m_c^2}{g_s^2} &=& 2 \epsilon_1^1 \epsilon_2^1 m_c \sqrt{\hat{s}} \chi + i \frac{(\epsilon_1^2 \epsilon_2^1 - \epsilon_1^1 \epsilon_2^2)M\sqrt{(k_{1T} - k_{2T})^2+\hat{s})(k_{1T} + k_{2T})^2+\hat{s})}}{\sqrt{\hat{s}}} \nonumber \\ 
&-& \frac{m_c}{\sqrt{\hat{s}}}\Big[ (\psi-\hat{s}) \epsilon_1^3 \epsilon_2^0 + (\psi+\hat{s}) \epsilon_1^0 \epsilon_2^3 + (\epsilon_1^0 \epsilon_2^0 - \epsilon_1^1 \epsilon_2^1 - \epsilon_1^2 \epsilon_2^2 + \epsilon_1^3 \epsilon_2^3) \sqrt{(k_{1T} - k_{2T})^2+\hat{s})(k_{1T} + k_{2T})^2+\hat{s})} \Big] \cos\theta^\prime \nonumber \\
&-& 2 m_c \sqrt{\hat{s}} \chi (\epsilon_1^1 \epsilon_2^1 - \epsilon_1^3 \epsilon_2^3 ) \cos^2\theta^\prime \nonumber \\
&-&\frac{m_c}{\sqrt{\hat{s}}}\Big[ (\psi-\hat{s}) \epsilon_1^1 \epsilon_2^0 + (\psi+\hat{s}) \epsilon_1^0 \epsilon_2^1 + (\epsilon_1^3 \epsilon_2^1 + \epsilon_1^1 \epsilon_2^3) \sqrt{(k_{1T} - k_{2T})^2+\hat{s})(k_{1T} + k_{2T})^2+\hat{s})} \Big] \sin\theta^\prime \nonumber \\
&+& 2 m_c \sqrt{\hat{s}} \chi (\epsilon_1^3 \epsilon_2^1 + \epsilon_1^1 \epsilon_2^3) \sin\theta^\prime \cos\theta^\prime \;,
\end{eqnarray}
\end{widetext}
where $\chi = \sqrt{1-4m_c^2/\hat{s}}$. The final state total spin is determined from the heavy quarks helicities. Two helicity combinations that result in $S_{z^\prime}=0$ are added and normalized to give the contribution to the spin triplet state ($S=1$) in Eqs. (\ref{s_triplet}), (\ref{t_triplet}), and (\ref{u_triplet}).

In this primed frame, to extract the projection on a state with orbital-angular-momentum quantum number $L$, we obtain the corresponding Legendre component $\mathcal{A}_L$ in the amplitudes by
\begin{eqnarray}
\mathcal{A}_{L=0} &=& \frac{1}{2} \int_{-1}^{1} dx \mathcal{A}(x=\cos \theta^\prime) \;, \\
\mathcal{A}_{L=1} &=& \frac{3}{2} \int_{-1}^{1} dx \; x \mathcal{A}(x=\cos \theta^\prime) \;.
\end{eqnarray}

\begin{figure*}
\centering
\begin{minipage}[ht]{0.97\columnwidth}
\centering
\includegraphics[width=\columnwidth]{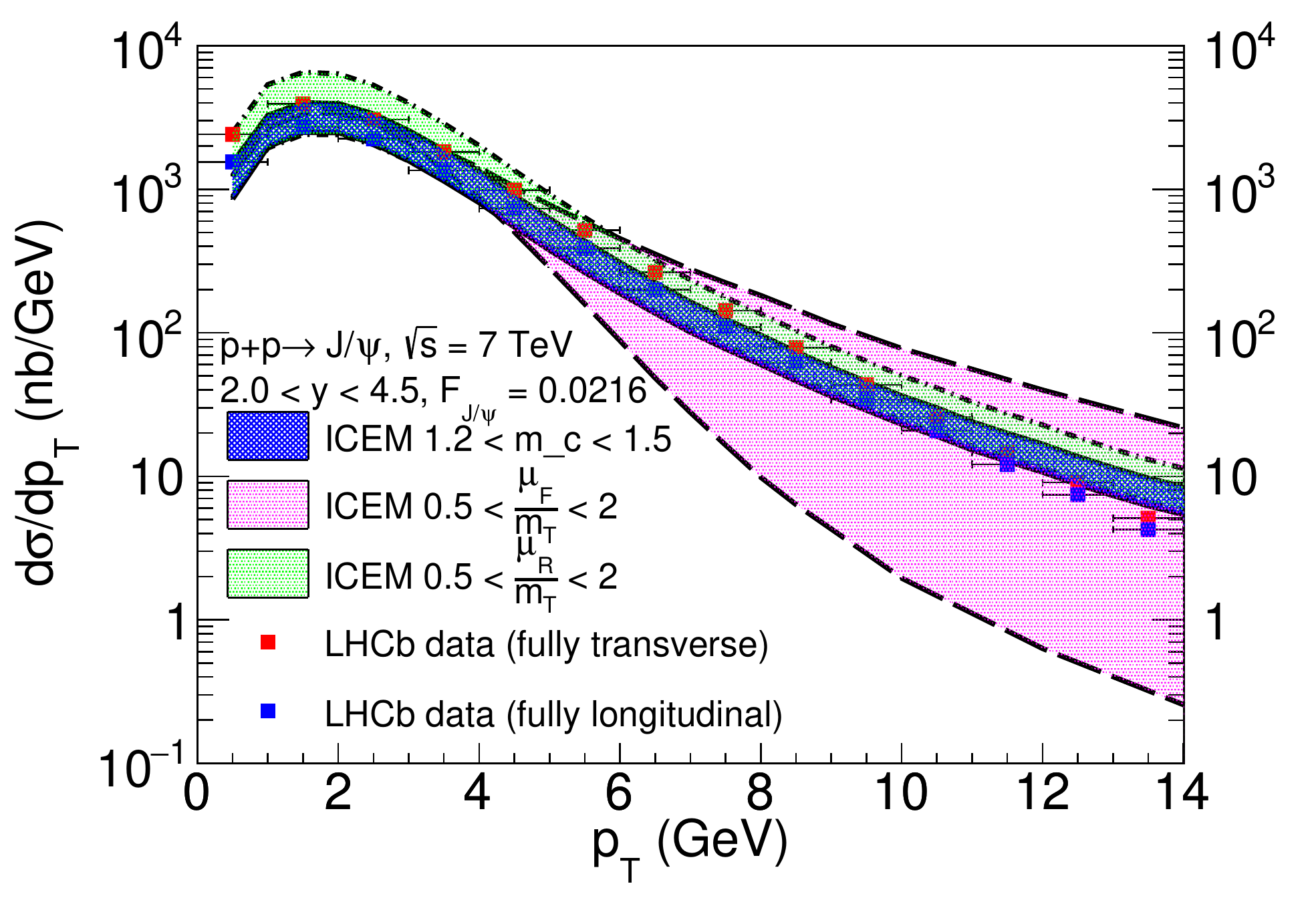}
\caption{The $p_T$ dependence of inclusive $J/\psi$ production at $\sqrt{s} = 7$~TeV in the ICEM obtained by varying the renormalization scale (blue solid), the factorization scale in the range $0.5<\mu_F/m_T<2$ (magenta dashed), and the renormalization scale in the range $0.5<\mu_R/m_T<2$ (green dot-dashed). The LHCb data  \cite{Aaij:2011jh} assuming the $J/\psi$ polarization is totally transverse, $\lambda_\vartheta = +1$ (red), and totally longitudinal, $\lambda_\vartheta = -1$ (blue), are shown. The LHCb data assuming $\lambda_\vartheta = 0$ lie between the red and blue points and are not shown.} \label{LHCB_1S_mass_and_doublemuF}
\end{minipage}%
\hspace{1cm}%
\begin{minipage}[ht]{0.97\columnwidth}
\centering
\includegraphics[width=\columnwidth]{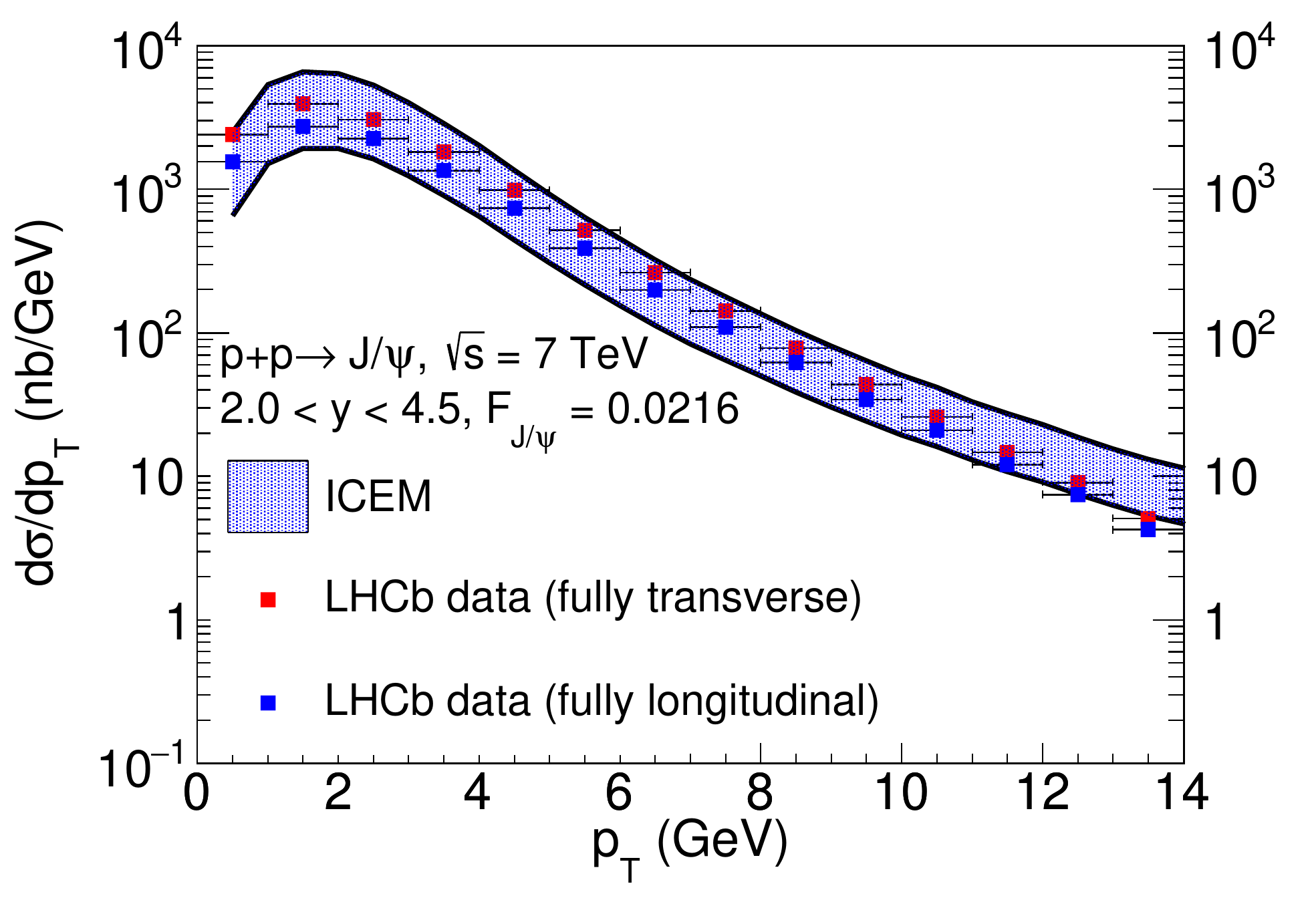}
\caption{The $p_T$ dependence of inclusive $J/\psi$ production at $\sqrt{s} = 7$~TeV in the ICEM with combined mass and renormalization scale uncertainties. The LHCb data  \cite{Aaij:2011jh} are shown as in Fig.~\ref{LHCB_1S_mass_and_doublemuF}. The LHCb data assuming $\lambda_\vartheta = 0$ are not shown.} \label{LHCb_1S_pt}
\end{minipage}
\end{figure*}

Having obtained the amplitudes for $S=1$ with $S_{z^\prime} = 0,\pm 1$, and $L=0,1$ with $L_{z^\prime} = 0$, we calculate the amplitudes for $J=0,1,2$. The amplitudes for $J=1$, found by adding $S=1$ and $L=0$, are
\begin{eqnarray}
\mathcal{A}_{J=1,J_{z^\prime}=\pm1} &=& \mathcal{A}_{L=0,L_{z^\prime}=0;S=1,S_{z^\prime}=\pm 1} \;, \\
\mathcal{A}_{J=1,J_{z^\prime}=0} &=& \mathcal{A}_{L=0,L_{z^\prime}=0;S=1,S_{z^\prime}=0} \;.
\end{eqnarray}
Employing angular momentum algebra, the amplitudes for $J=0,1,2$, obtained by adding $S=1$ and $L=1$, are
\begin{eqnarray}
\mathcal{A}_{J=0,J_{z^\prime}=0} &=& -\sqrt{\frac{1}{3}} \mathcal{A}_{L=1,L_{z^\prime}=0;S=1,S_{z^\prime}=0} \;, \\
\mathcal{A}_{J=1,J_{z^\prime}=\pm1} &=& \mp \frac{1}{\sqrt{2}} \mathcal{A}_{L=1,L_{z^\prime}=0;S=1,S_{z^\prime}=\pm 1} \;, \\
\mathcal{A}_{J=1,J_{z^\prime}=0} &=& 0 \;, \\
\mathcal{A}_{J=2,J_{z^\prime}=\pm2} &=& 0 \;, \label{J_z=pm2} \\
\mathcal{A}_{J=2,J_{z^\prime}=\pm1} &=& \frac{1}{\sqrt{2}} \mathcal{A}_{L=1,L_z^\prime=0;S=1,S_{z^\prime}=\pm1} \;, \\
\mathcal{A}_{J=2,J_{z^\prime}=0} &=& \sqrt{\frac{2}{3}} \mathcal{A}_{L=1,L_{z^\prime}=0;S=1,S_{z^\prime}=0} \;.
\end{eqnarray}

\begin{table}
\caption{\label{states}The mass $M_\mathcal{Q}$, the feed-down contribution ratio $c_\mathcal{Q}$, and the squared feed-down transition Clebsch-Gordan coefficients $S_\mathcal{Q}^{J_z}$ for all quarkonium states contributing to prompt $J/\psi$ production.}
\begin{ruledtabular}
\begin{tabular}{ccccc}
$\mathcal{Q}$ & $M_\mathcal{Q}$ (GeV) & $c_{\mathcal{Q}}$ & $S_\mathcal{Q}^{J_z=0}$ & $S_\mathcal{Q}^{J_z=\pm1}$\\
\hline
$J/\psi$ & 3.10 & 0.62 & 1 & 0 \\
$\psi$(2S) & 3.69 & 0.08 & 1 & 0 \\
$\chi_{c1}$(1P) & 3.51 & 0.16 & 0 & 1/2 \\
$\chi_{c2}$(1P) & 3.56 & 0.14 & 2/3 & 1/2 \\
\end{tabular}
\end{ruledtabular}
\end{table}

Using a Wigner representation of the inverse rotation defined in Eq. (\ref{forward_rotation}),
\begin{eqnarray}
\mathcal{D}^J_{J_z,J_{z^\prime}} = \bra{J,J_z} \mathcal{R}(0,-\beta,-\gamma) \ket{J,J_{z^\prime}} \; ,
\end{eqnarray}
the amplitudes sorted by final state $J$ and $J_{z^\prime}$ are then rotated back into the helicity frame:
\begin{eqnarray}
\mathcal{A}_{J,J_z} = \sum_{J_z^\prime = -J}^{J} \mathcal{D}^J_{J_z,J_{z^\prime}} \mathcal{A}_{J,J_z^\prime} \;.
\end{eqnarray}
Next, the amplitudes sorted by final state $J$ and $J_{z}$ are squared for calculations in the helicity frame. For calculations in the other frames, the unsquared amplitudes can be further rotated to the Collins-Soper (CS) or the Gottfried-Jackson (GJ) frame. In the CS frame, the $z$-axis is defined as the angle bisector of the angle between one proton beam and the opposite of the other proton beam. In the GJ frame, the $z$-axis is defined as the direction of the momentum of one of the two colliding proton beams.

The squared matrix elements, $|\mathcal{M}|^2$, are calculated for each $J$, $J_z$ combination. The color factors, $C$, are calculated from the SU(3) color algebra and are independent of final state angular momentum \cite{Cvitanovic:1976am}. They are
\begin{eqnarray}
|C_{gg,\hat{s}}|^2 = 12 \;, \nonumber \\ 
|C_{gg,\hat{t}}|^2 = \frac{16}{3} \;, |C_{gg,\hat{u}}|^2 = \frac{16}{3} \; .
\end{eqnarray}
\begin{eqnarray}
C_{gg,\hat{s}}^*C_{gg,\hat{t}} = +6 \;, C_{gg,\hat{s}}^*C_{gg,\hat{u}} = -6 \;, \nonumber \\
C_{gg,\hat{t}}^*C_{gg,\hat{u}} = -\frac{2}{3} \;.
\end{eqnarray}
Finally, the total squared amplitudes for a given $J,J_z$ state,
\begin{eqnarray}
|\mathcal{M}_{gg}^{J,J_z}|^2 &=&  |C_{gg,\hat{s}}|^2 |\mathcal{A}_{gg,\hat{s}}|^2 + |C_{gg,\hat{t}}|^2 |\mathcal{A}_{gg,\hat{t}}|^2 \nonumber \\
&+& |C_{gg,\hat{u}}|^2 |\mathcal{A}_{gg,\hat{u}}|^2 + 2  C_{gg,\hat{s}}^*C_{gg,\hat{t}} \mathcal{A}_{gg,\hat{s}}^*\mathcal{A}_{gg,\hat{t}} \nonumber \\
&+& 2 C_{gg,\hat{s}}^*C_{gg,\hat{u}} \mathcal{A}_{gg,\hat{s}}^*\mathcal{A}_{gg,\hat{u}} \nonumber \\ 
&+& 2 C_{gg,\hat{t}}^*C_{gg,\hat{u}} \mathcal{A}_{gg,\hat{t}}^*\mathcal{A}_{gg,\hat{u}} \;,
\end{eqnarray}
are then employed to calculate the partonic cross sections by integrating over solid angle
\begin{eqnarray}
\hat{\sigma}^{J,J_z} &=& \int d\Omega \Big( \frac{1}{8\pi} \Big)^2 |\mathcal{M}^{J,J_z}|^2 \\
&\times& \frac{2\chi}{\sqrt{((k_{1T} - k_{2T})^2+\hat{s})((k_{1T} + k_{2T})^2+\hat{s})}} \nonumber \;.
\end{eqnarray}

The sum of the polarized partonic cross sections results for each final state total angular momentum $J$, is equal to the unpolarized partonic cross section,
\begin{eqnarray}
\hat{\sigma}_{\rm unpol}=\sum_{J_z=-J}^{J_z=+J} \hat{\sigma}^{J,J_z} \;.
\end{eqnarray}

Having computed the polarized $Q\bar{Q}$ production cross section at the parton level, we then convolute the partonic cross sections with the unintegrated parton distribution functions (uPDFs) to obtain the hadron-level cross section $\sigma$ as a function of $p_T$ using Eq.~(\ref{pt_rap_cut}) or (\ref{pt_xf_cut}) and as a function of $y$ using Eq.~(\ref{y_pt_cut}). The quarkonium masses which appear as the lower limit of the $Q\bar{Q}$ invariant mass are listed in Table~\ref{states}. We employ the ccfm-JH-2013-set1 \cite{Hautmann:2013tba} uPDFs in this calculation. 


\section{Polarization of prompt $J/\psi$}
We assume that the angular momentum of each directly-produced quarkonium state is unchanged by the transition from the parton level to the hadron level, consistent with the CEM expectation that the linear momentum is unchanged by hadronization. This is similar to the assumption made in NRQCD that once a $c \bar c$ is produced in a given spin state, it retains that spin state when it becomes a $J/\psi$.

We calculate the $J_z=0,\pm1$ to unpolarized ratios for each directly produced quarkonium state $\mathcal{Q}$ that has a contribution to prompt $J/\psi$ production: $J/\psi$, $\psi$(2S), $\chi_{c1}$(1P) and $\chi_{c2}$(1P). These ratios, $R_{\mathcal{Q}}^{J_z}$, are then independent of $F_{\mathcal{Q}}$. We assume the feed-down production of $J/\psi$ from the higher mass bound states follows the angular momentum algebra. Their contributions to the $J_z=0$ to unpolarized ratios of prompt $J/\psi$ are added and weighed by the feed-down contribution ratios $c_{\mathcal{Q}}$ \cite{Digal:2001ue},

\begin{eqnarray}
\label{mix_psi}
R_{J/\psi}^{J_z=0} &=& \sum_{\mathcal{Q},J_z} c_{\mathcal{Q}} S_{\mathcal{Q}}^{J_z} R_{\mathcal{Q}}^{J_z} \;,
\end{eqnarray}
where $S_{\mathcal{Q}}^{J_z}$ is the transition probability from a given state $\mathcal{Q}$ produced in a $J_z$ state to a $J/\psi$ with $J_z=0$ in a single decay. We assume two pions are emitted for S state feed down, $\psi{\rm (2S)}\rightarrow J/\psi \pi \pi$, and a photon is emitted for a P state feed down, $\chi_{c}\rightarrow J\psi \gamma$. $S_{\mathcal{Q}}^{J_z}$ is then 1 (if $J_z=0$) or 0 (if $J_z=1$) for $\mathcal{Q}=\psi$(2S) since the transition, $\psi{\rm (2S)} \rightarrow J/\psi \pi \pi$, does not change the angular momentum of the quarkonium state. For directly produced $J/\psi$, $S_{\mathcal{Q}}^{J_z}$ is 1 for $J_z=0$ and 0 for $J_z=1$. The $S_{\mathcal{Q}}^{J_z}$ for the $\chi$ states are the squares of the Clebsch-Gordan coefficients for the feed-down production via $\chi \rightarrow J/\psi+\gamma$.  The values of $M_\mathcal{Q}$, $c_\mathcal{Q}$, and $S_\mathcal{Q}^{J_z}$ for all quarkonium states contributing to prompt $J/\psi$ production are collected in Table~\ref{states}.

\begin{figure*}
\centering
\begin{minipage}[ht]{0.97\columnwidth}
\centering
\includegraphics[width=\columnwidth]{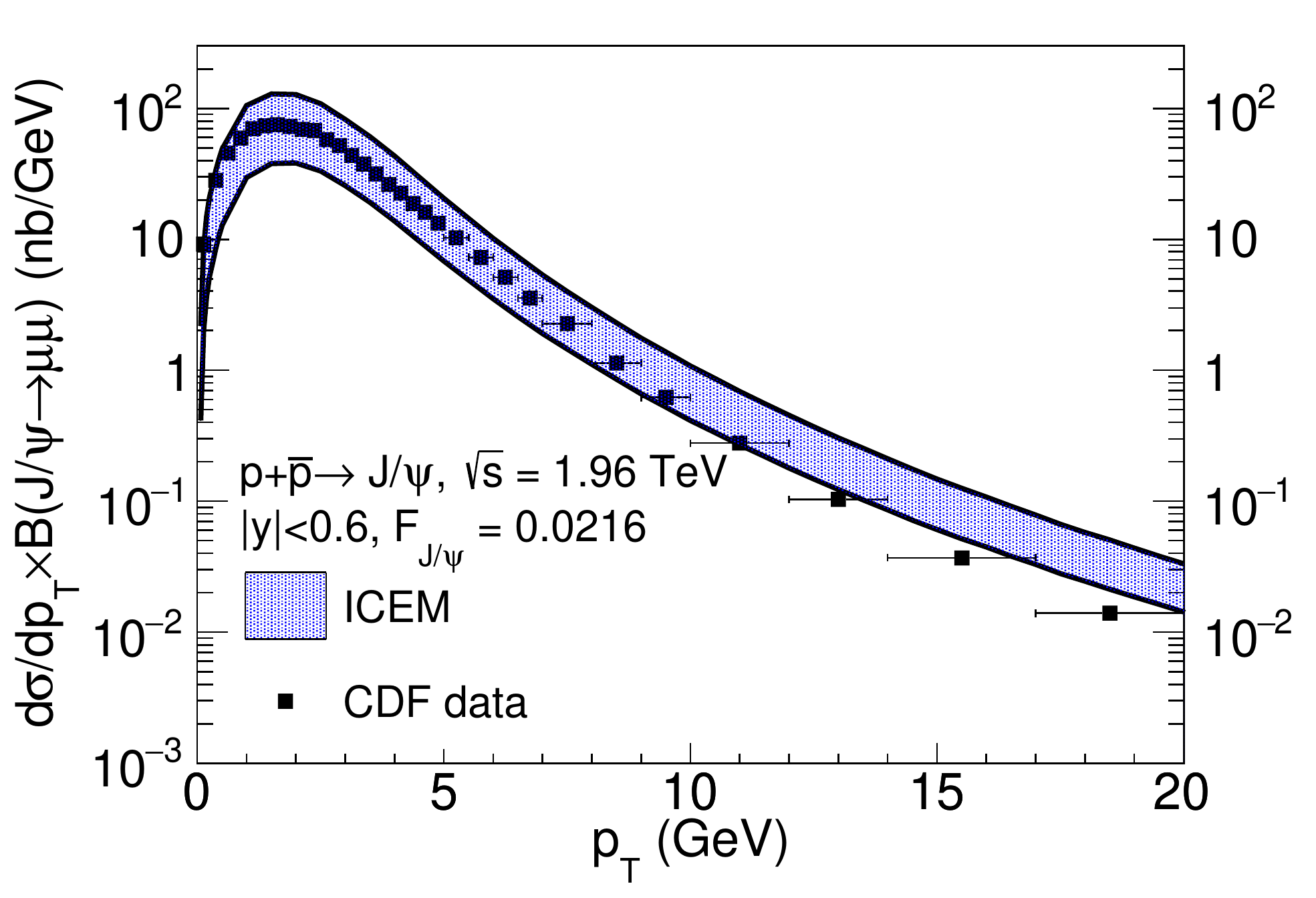}
\caption{The $p_T$ dependence of inclusive $J/\psi$ production at $\sqrt{s} = 1.96$~TeV in the ICEM. The combined mass and renormalization scale uncertainties are shown in the band and compared to the CDF data \cite{Acosta:2004yw}.} \label{CDF_1S_pt}
\end{minipage}%
\hspace{1cm}%
\begin{minipage}[ht]{0.97\columnwidth}
\centering
\includegraphics[width=\columnwidth]{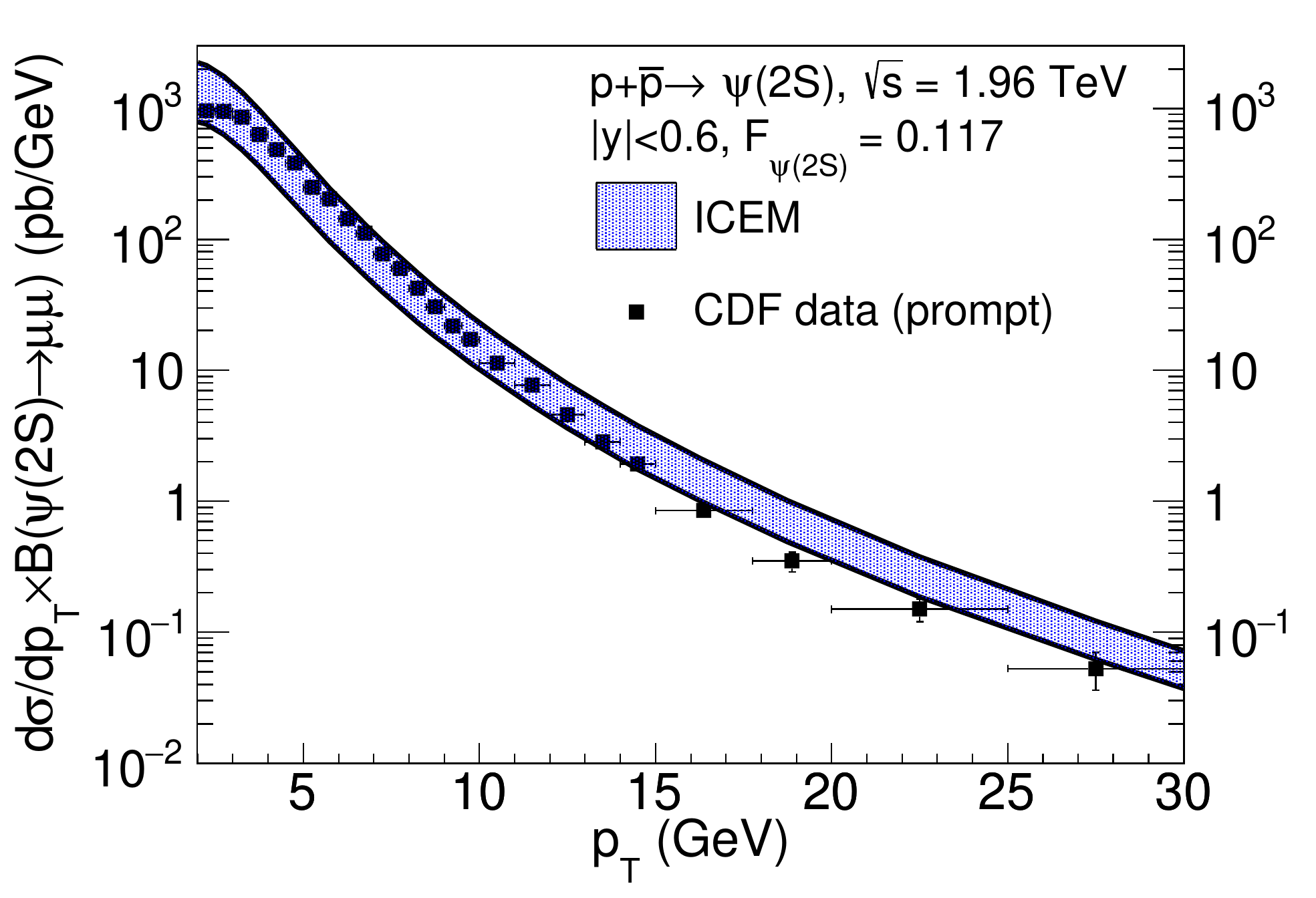}
\caption{The $p_T$ dependence of direct $\psi$(2S) production at $\sqrt{s} = 1.96$~TeV in the ICEM. The combined mass and renormalization scale uncertainties are shown in the band and compared to the CDF data for prompt $\psi$(2S) \cite{Aaltonen:2009dm}.} \label{CDF_2S_pt}
\end{minipage}
\end{figure*}

Finally, the $J_z=0$ to the unpolarized ratio for prompt $J/\psi$ is converted into the polarization parameter $\lambda_\vartheta$ \cite{Faccioli:2010kd},
\begin{eqnarray}
\label{s_state_lambda}
\lambda_{\vartheta} &=& \frac{1-3R^{J_z=0}}{1+R^{J_z=0}} \;,
\end{eqnarray}
where $-1<\lambda_\vartheta<1$. If $\lambda_\vartheta = -1$, $J/\psi$ production is totally longitudinal, $\lambda_\vartheta= 0$ refers to unpolarized production, and for $\lambda_\vartheta=+1$, production is totally transverse.


\section{Results}

Although the matrix element in this calculation is LO in $\alpha_s$, by convoluting the polarized partonic cross sections with the transverse momentum dependent uPDFs using the $k_T$-factorization approach, we can calculate the yield as well as the polarization parameter $\lambda_\vartheta$ as a function of $p_T$. The full NLO polarization including  $q\bar{q}$ and $(q+\bar{q})g$ contributions, requiring us to go to $\mathcal{O}(\alpha_s^3)$, will be discussed in a future publication.

The traditional CEM can describe the unpolarized yield of charm and $J/\psi$ production at both LO and NLO assuming collinear factorization \cite{Vogt:1995zf,NVF}. The ICEM can also describe the $\psi$(2S) to $J/\psi$ ratio at NLO while, in the traditional CEM, this ratio is independent of $p_T$ \cite{Ma:2016exq}. Since this is the first calculation in the ICEM using the $k_T$-factorization approach, it is important to check if the unpolarized yield is also in agreement with the data.

In the remainder of this section, we first present how our approach describes the transverse momentum and rapidity distribution of the charmonium states in collider experiments. We then discuss the transverse momentum and rapidity dependence of the polarization parameter $\lambda_\vartheta$ for prompt $J/\psi$ production and direct production of quarkonium states that contribute to the feed-down production. We compare our results to the polarization measured in fixed-target experiments as well as collider experiments in the helicity, Collins-Soper, and Gottfried-Jackson frames to discuss the frame dependence of the polarization parameter. Finally, we discuss the sensitivity of our results to the factorization and renormalization scales, the weight of each diagram, and the feed-down ratios considered. In our calculations, we construct the uncertainty bands by varying the charm quark mass, around its base value of 1.27~GeV in the interval $1.2 < m_c < 1.5$~GeV, and the renormalization scale around its base value of $m_T$ in the interval $0.5 < \mu_R/m_T < 2$ while keeping the factorization scale fixed at $\mu_F=m_T$. The total uncertainty band is constructed by adding the two uncertainties in quadrature.

\begin{figure*}
\centering
\begin{minipage}[ht]{0.6873\columnwidth}
\centering
\includegraphics[width=\columnwidth]{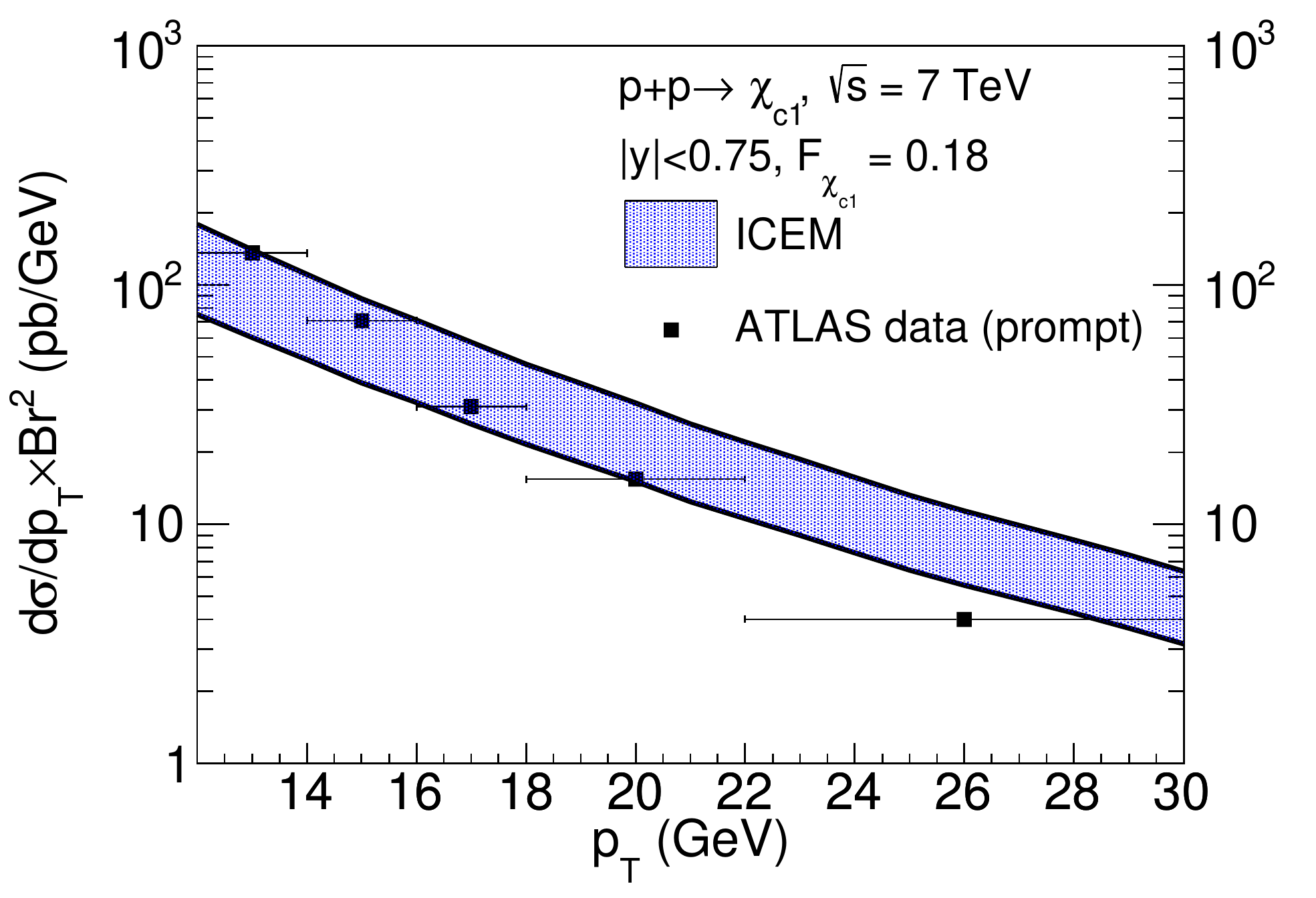}
\end{minipage}%
\begin{minipage}[ht]{0.6873\columnwidth}
\centering
\includegraphics[width=\columnwidth]{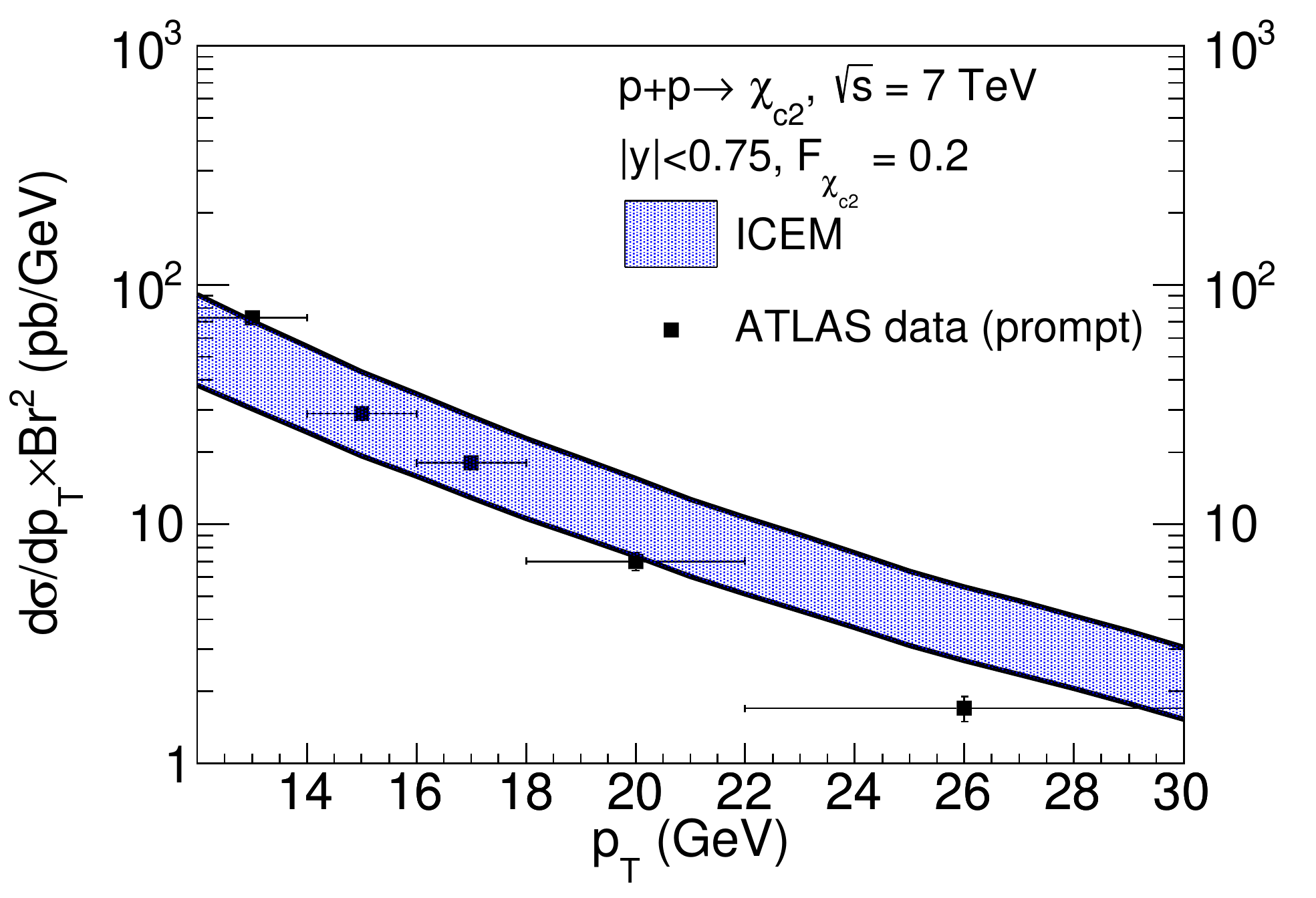}
\end{minipage}
\begin{minipage}[ht]{0.6873\columnwidth}
\centering
\includegraphics[width=\columnwidth]{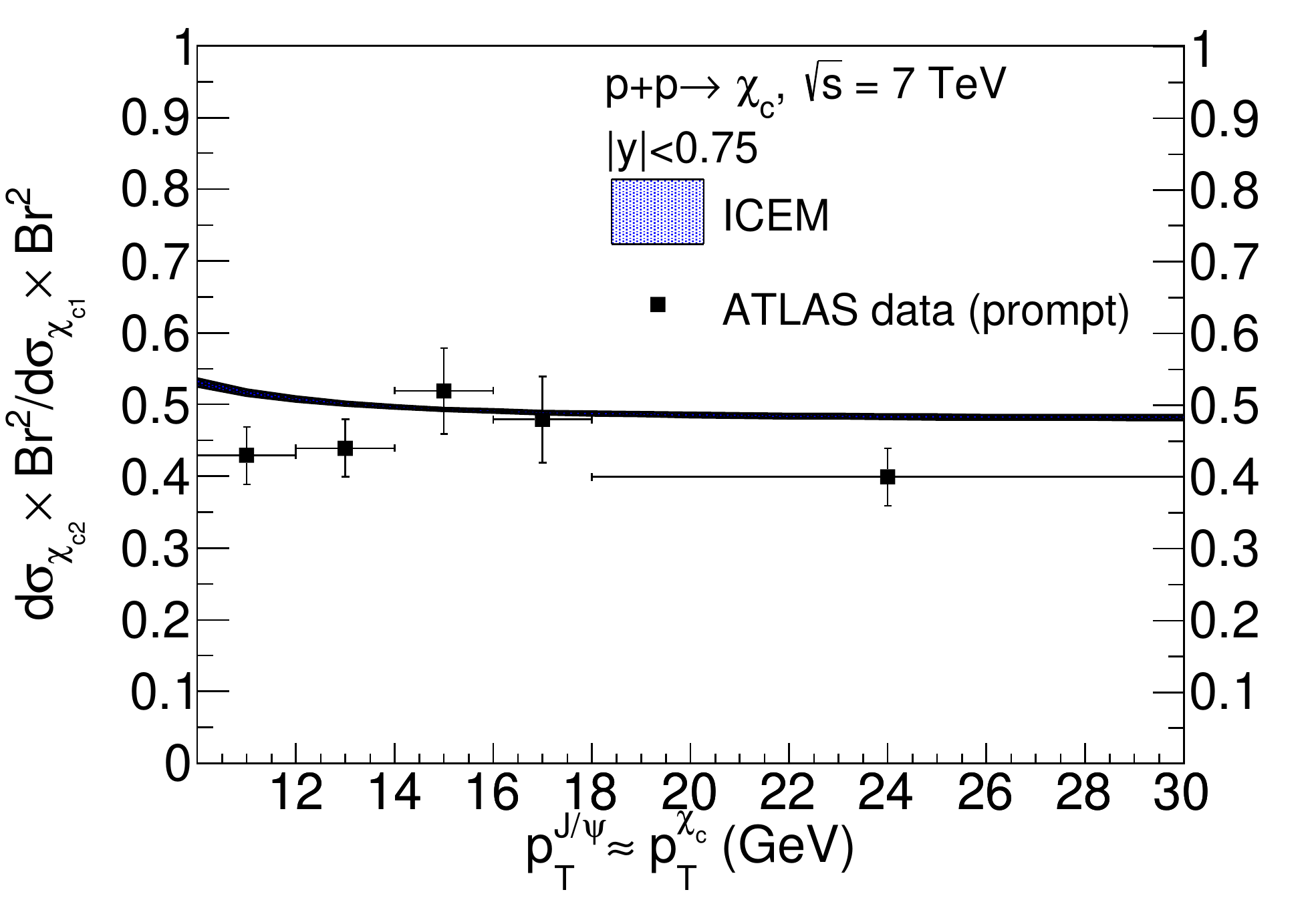}
\end{minipage}
\\
\centering
\begin{minipage}[ht]{0.6873\columnwidth}
\centering
\includegraphics[width=\columnwidth]{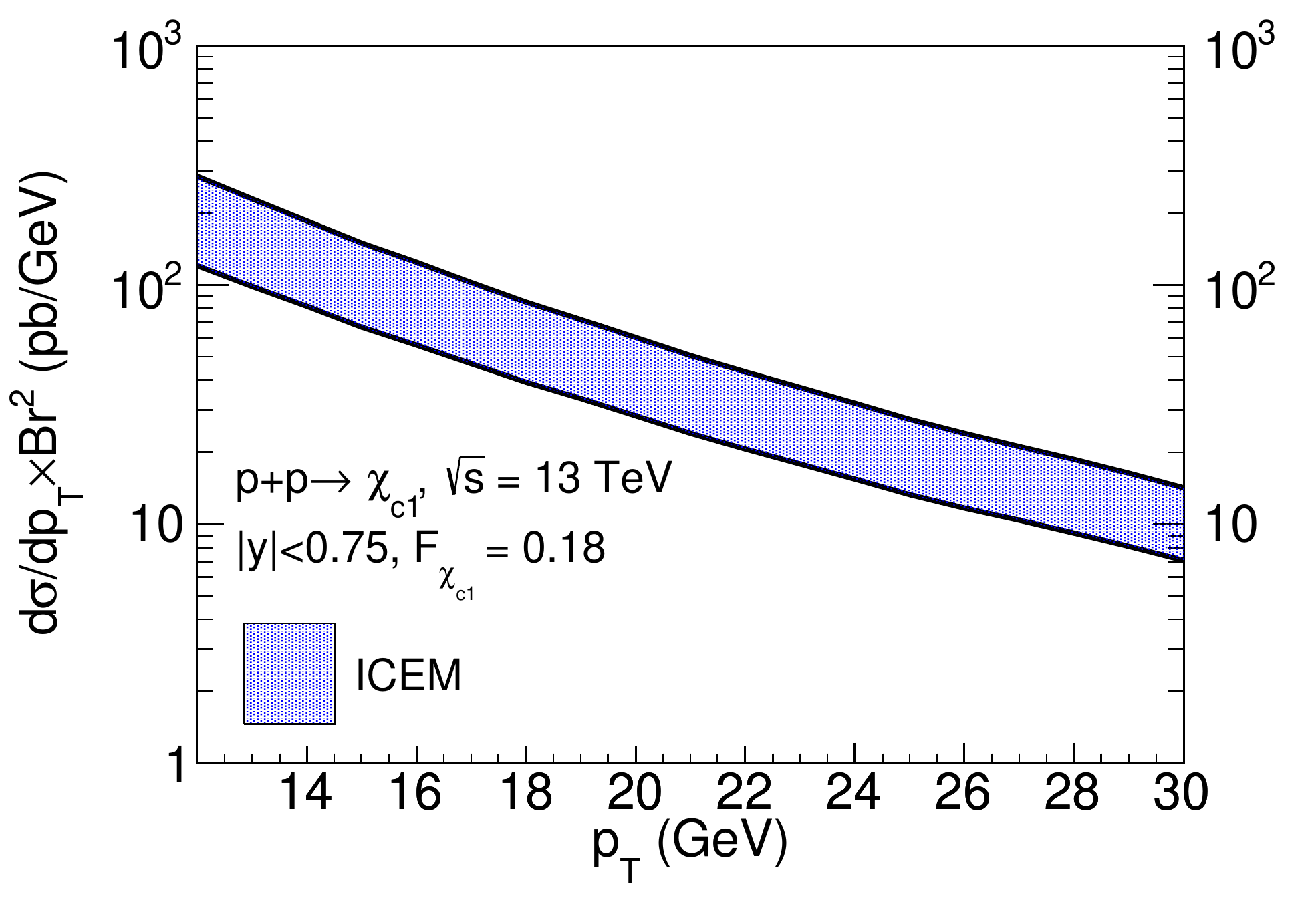}
\end{minipage}%
\begin{minipage}[ht]{0.6873\columnwidth}
\centering
\includegraphics[width=\columnwidth]{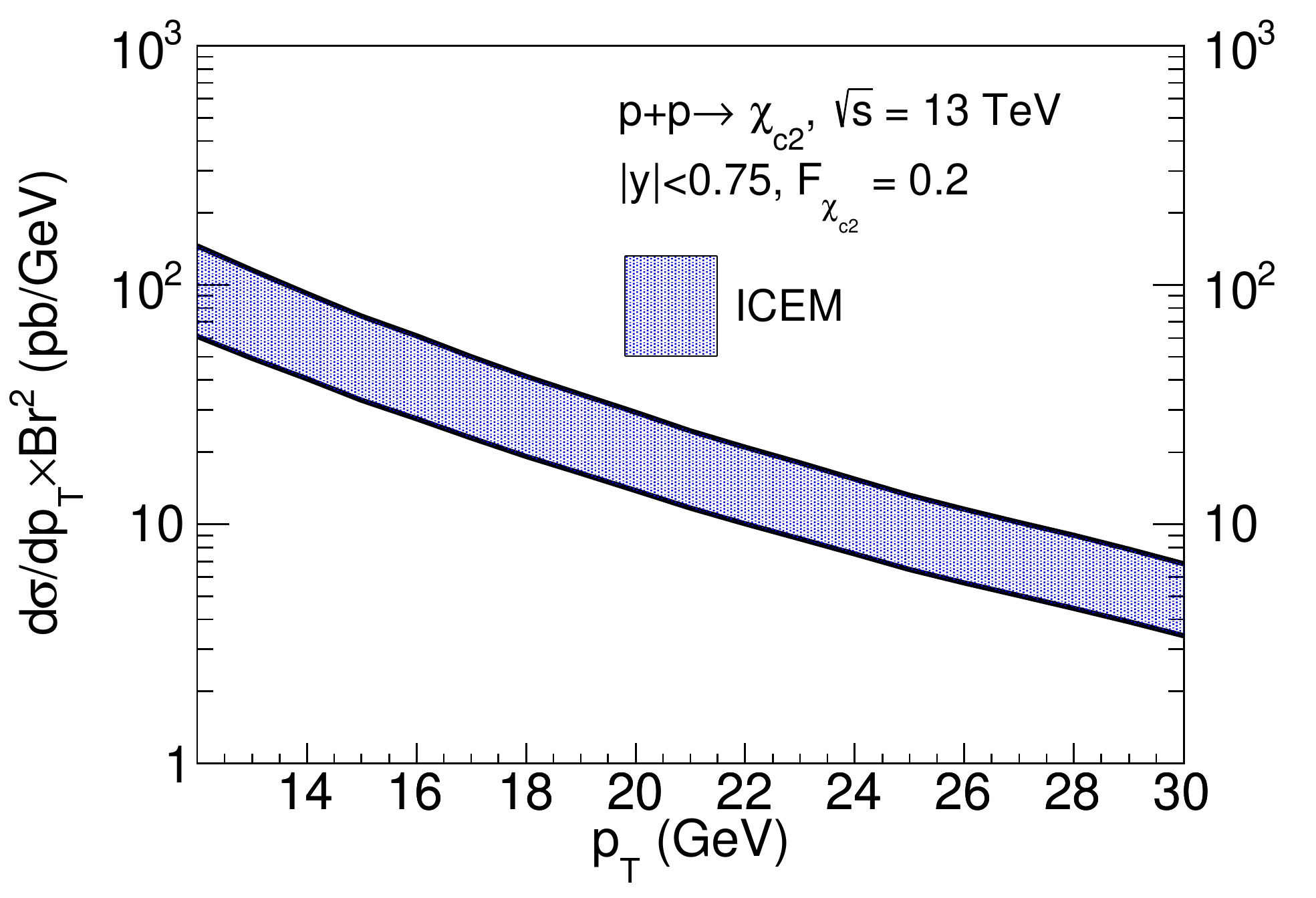}
\end{minipage}
\begin{minipage}[ht]{0.6873\columnwidth}
\centering
\includegraphics[width=\columnwidth]{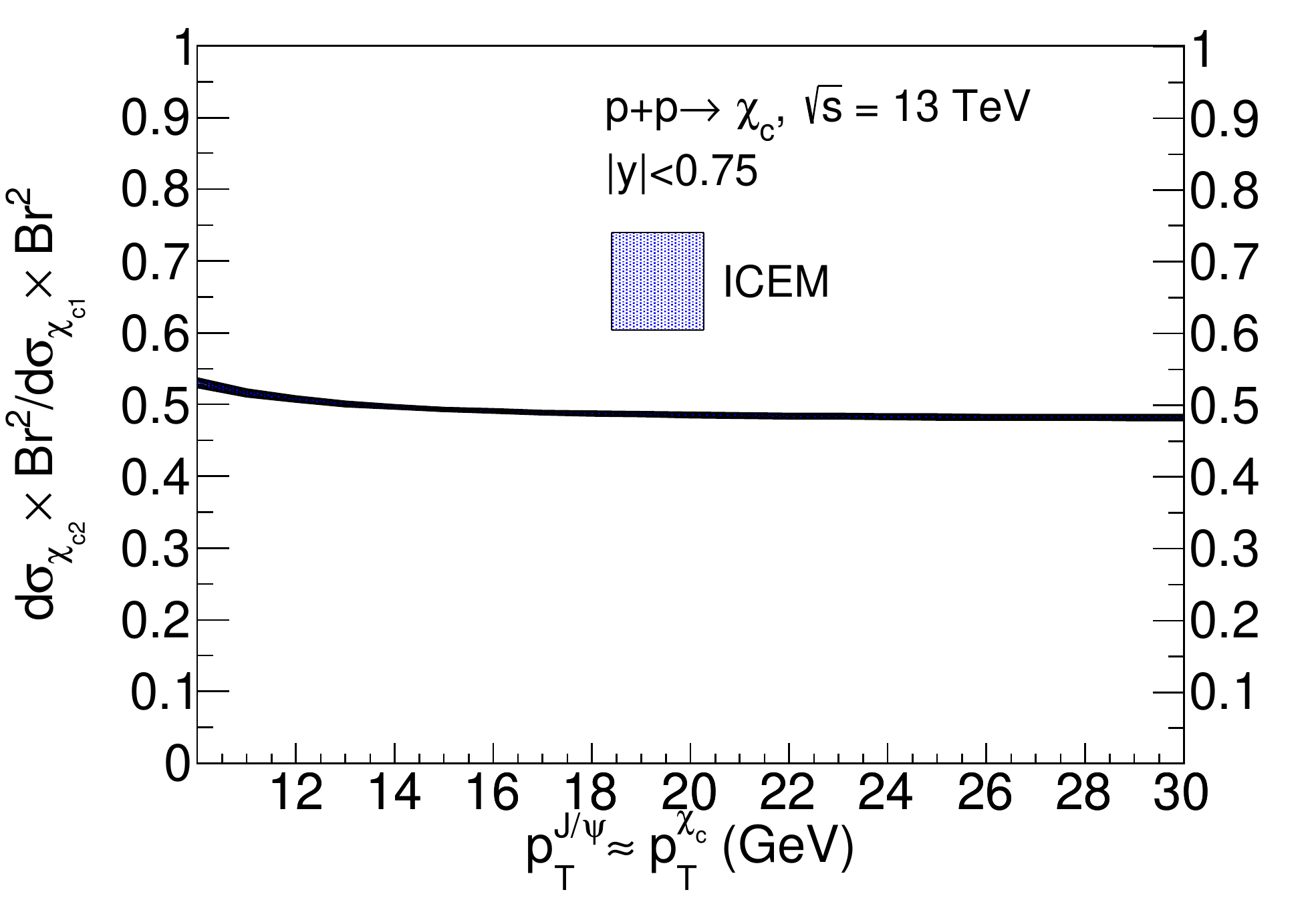}
\end{minipage}
\caption{The direct $\chi_{c1}$ (left) and $\chi_{c2}$ (middle) $p_T$ dependence multiplied by the branching ratios for $\chi_c \rightarrow J/\psi \gamma$ and for $J/\psi \rightarrow \mu^+\mu^-$, and the ratio of $\chi_{c2}$ to $\chi_{c1}$ (right) at $\sqrt{s} = 7$~TeV (top panels) and at $\sqrt{s}=13$~TeV (bottom panels) in the ICEM with combined mass and renormalization scale uncertainties. The ATLAS data for prompt $\chi_{c}$ production are also shown \cite{ATLAS:2014ala}.} \label{ATLAS_chic}
\end{figure*}

\subsection{Unpolarized charmonium production}

In this section, we present the $p_T$ and rapidity distributions of charmonium states in our approach. In the spirit of the traditional CEM, $F_\mathcal{Q}$ in Eq.(\ref{cem_sigma}) has to be independent of the projectile, target, and energy for each quarkonium state $\mathcal{Q}$. Even though the focus of this paper is on polarization, which is $F_\mathcal{Q}$ independent, the unpolarized yield in the ICEM using the $k_T$-factorization approach was not considered before. Therefore, it is important to first confirm that this approach can indeed describe the charmonium yields as a function of $p_T$ and rapidity before discussing polarization predictions. We first obtain $F_{J/\psi}$ and $F_{\psi{\rm (2S)}}$ by comparing our results with the experimental data measured by the LHCb Collaboration and the CDF Collaboration respectively. Using the same $F_{J/\psi}$ and $F_{\psi{\rm (2S)}}$, we compare our results with the experimental data measured at CDF and ALICE. We can only obtain $F_{\chi_{\rm c1}}$ and $F_{\chi_{\rm c2}}$ for the $\chi_c$ states by comparing the unpolarized yield with the data measured by the ATLAS Collaboration at $\sqrt{s}=7$~TeV because these are the only measurements. We instead give predictions of $\chi_{\rm c1}$ and $\chi_{\rm c1}$ production at $\sqrt{s}=13$~TeV. We also compare and predict the ratio of $\chi_{c2}$ to $\chi_{c1}$ at $\sqrt{s}=7$~TeV and $\sqrt{s}=13$~TeV. Note that we cannot expect that our LO values of $F_Q$ to be equal to those found for $J\psi$ and $\psi$(2S) in Ref.~\cite{Ma:2016exq}. Those calculations are NLO in the total cross section assuming collinear factorization, and include the $q\bar{q}$ and $(q+\bar{q})g$ channels where the contribution of the later is non-negligible.

\begin{figure}[hb]
\centering
\includegraphics[width=\columnwidth]{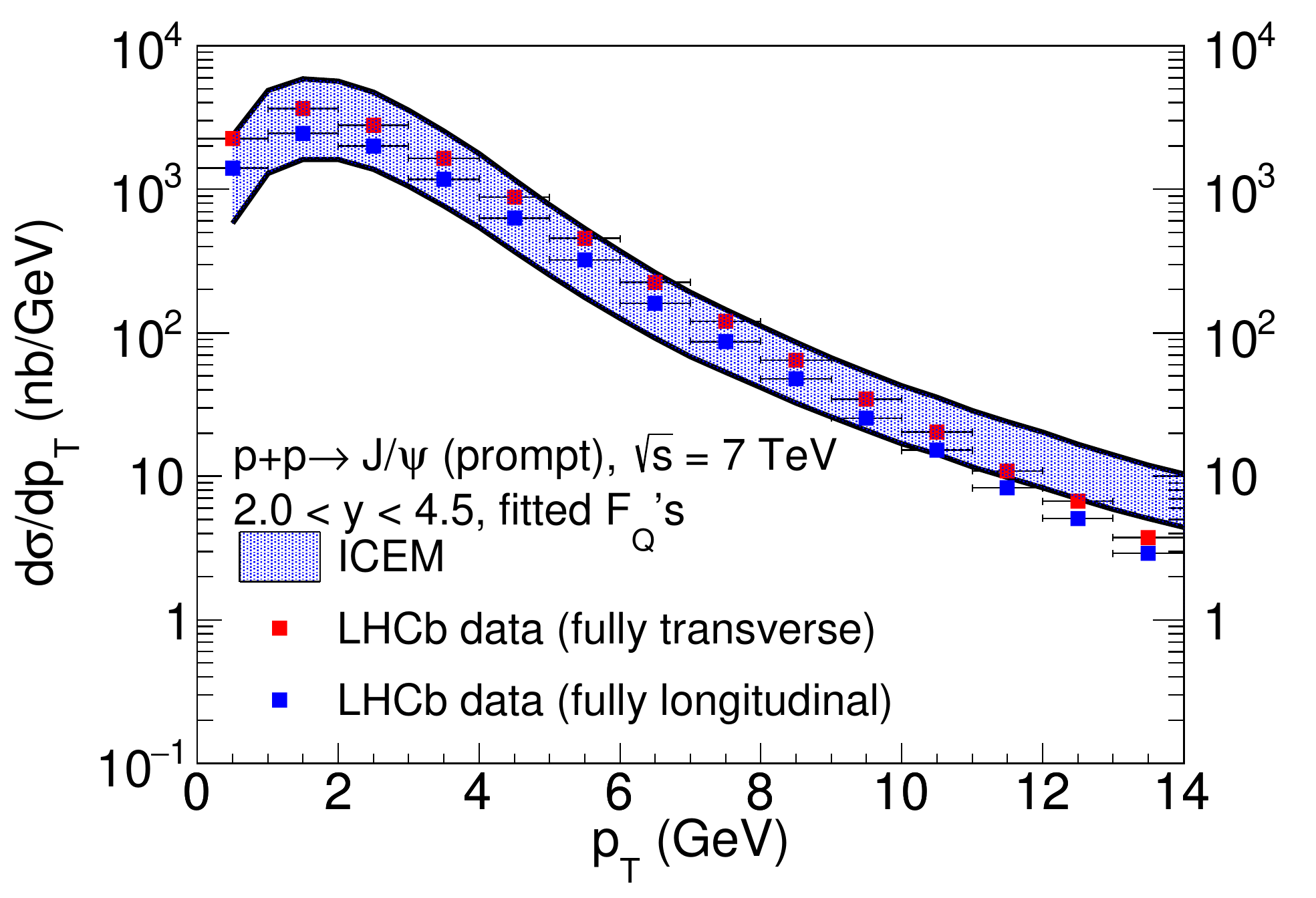}
\caption{The $p_T$ dependence of prompt $J/\psi$ production at $\sqrt{s} = 7$~TeV in the ICEM using fitted $F_\mathcal{Q}$'s with combined mass and renormalization scale uncertainties. The LHCb data  \cite{Aaij:2011jh} are shown as in Fig.~\ref{LHCB_1S_mass_and_doublemuF}. The LHCb data assuming $\lambda_\vartheta = 0$ are not shown.} \label{LHCb_1S_prompt}
\end{figure}


\begin{figure*}
\centering
\begin{minipage}[ht]{0.97\columnwidth}
\centering
\includegraphics[width=\columnwidth]{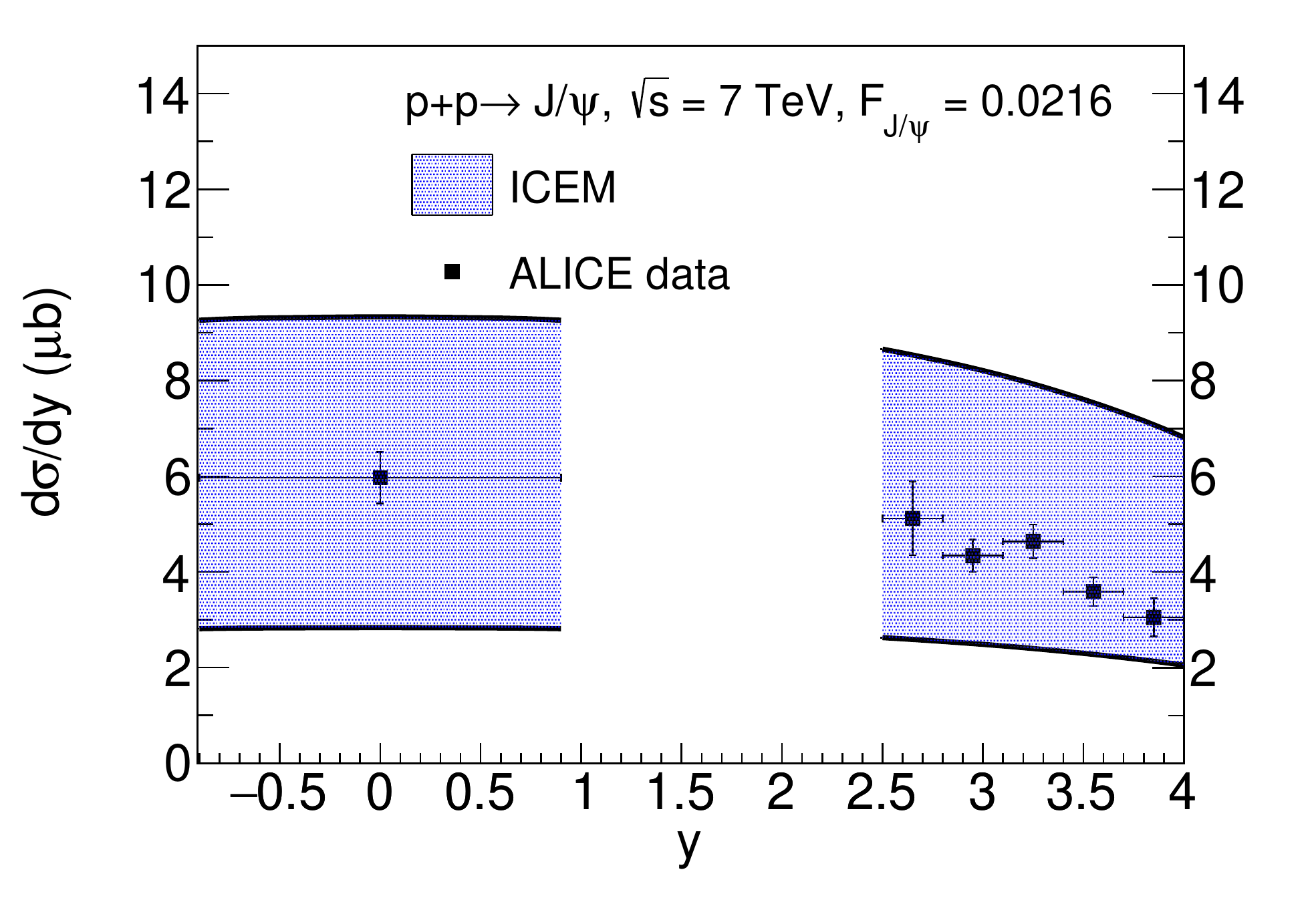}
\caption{The rapidity dependence of inclusive $J/\psi$ production at $\sqrt{s} = 7$~TeV in the ICEM. The combined mass and renormalization scale uncertainties are shown in the band and compared to the ALICE data \cite{Aamodt:2011gj}.} \label{ALICE_1S_rapidity}
\end{minipage}%
\hspace{1cm}%
\begin{minipage}[ht]{0.97\columnwidth}
\centering
\includegraphics[width=\columnwidth]{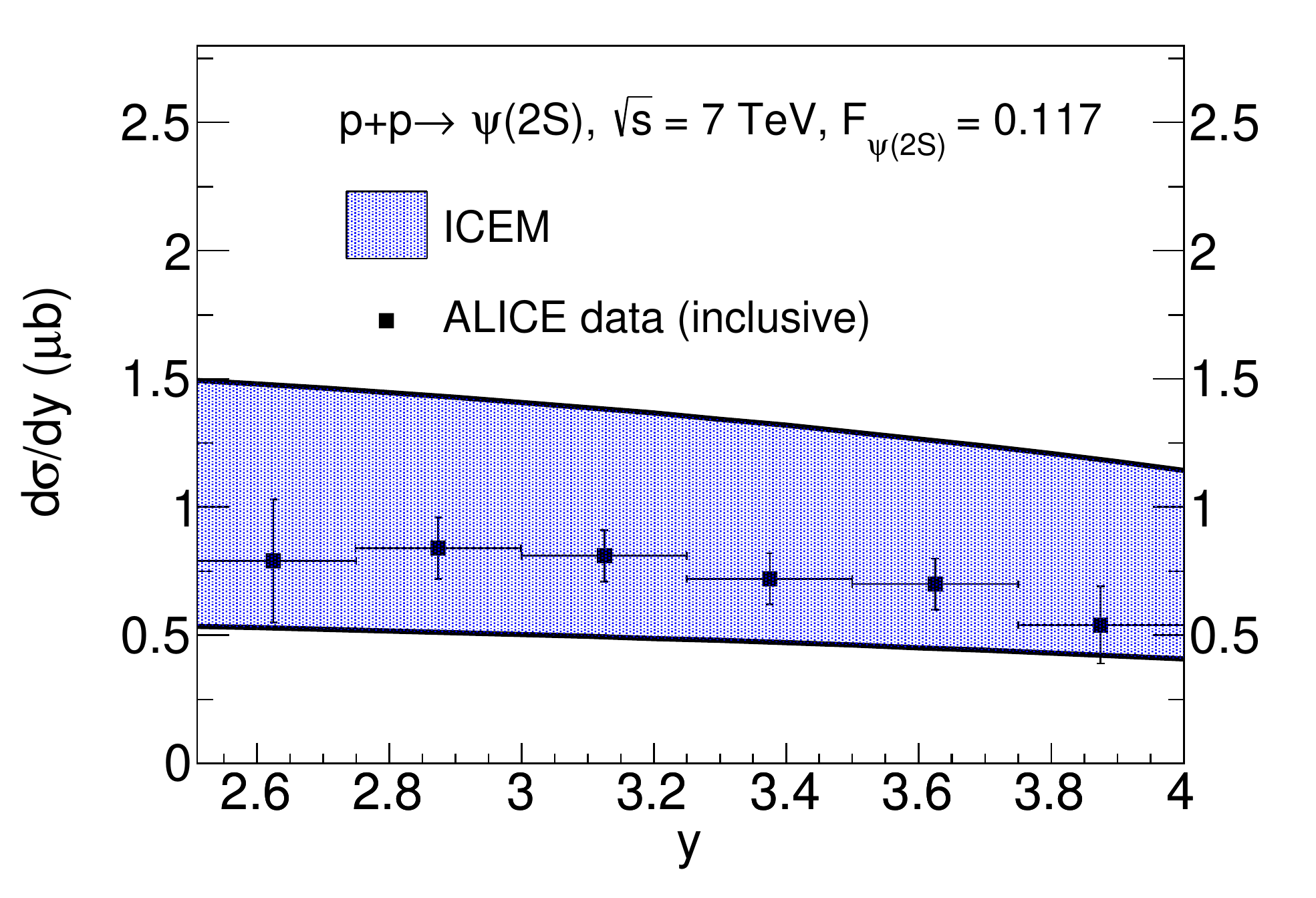}
\caption{The rapidity dependence of direct $\psi$(2S) production at $\sqrt{s} = 7$~TeV in the ICEM. The combined mass and renormalization scale uncertainty are shown in the band and compared to the ALICE data for inclusive $\psi$(2S) \cite{Abelev:2014qha}.} \label{ALICE_2S_rapidity}
\end{minipage}
\end{figure*}

\subsubsection{$J/\psi$ $p_T$ distribution}
\label{factorization}
We first discuss why we fix the factorization scale at $\mu_F=m_T$ instead of including a factor of two variation, as usual in most other approaches. In Fig.~\ref{LHCB_1S_mass_and_doublemuF}, we show the $p_T$ distributions of inclusive $J/\psi$ production at $\sqrt{s} = 7$~TeV found by fixing $m_c = 1.27$~GeV, and varying the factorization scale over the range $0.5 < \mu_F/m_T < 2$ and the renormalization scale over the range $0.5 < \mu_R/m_T < 2$ separately. We also fix $\mu_F/m_T=\mu_R/m_T=1$ and vary the charm quark mass over the range $1.2<m_c<1.5$~GeV. The direct production cross section is calculated using Eq.~(\ref{pt_rap_cut}) by integrating the pair invariant mass from $M_{J/\psi}$ to $2m_{D^0}$ ($m_{D^0}=1.86$~GeV) over the rapidity range $2.0 < y < 4.5$. We assume the direct production is a constant fraction, $0.62$ of the inclusive production \cite{Digal:2001ue}. We then are able to compare the inclusive $p_T$ distribution in the ICEM with the LHCb data \cite{Aaij:2011jh}. The result has a significant dependence on the factorization scale for $p_T>5$~GeV. This is because the uPDFs have a sharp cutoff for $k_T > \mu_F$ and are thus very sensitive to the chosen factorization scale. The yield varies more as $p_T$ approaches $m_T$ at high $p_T$. At low $p_T$, $m_T \sim M_\mathcal{Q}$ and the cross section is independent of the factorization scale since $k_T << \mu_F$. At moderate $p_T$, the variation with $\mu_F$ is similar to or smaller than that due to the charm quark mass. At $p_T\sim10$~GeV, $m_T \sim p_T$. Thus the lower limit on the factorization scale, $m_T/2$, is on the order of $k_T$ and the yield drops off at this cutoff limit, while the upper limit on the factorization scale, $2m_T$, is still greater than $k_T$, enhancing the yield. Since at LO, only the $Q\bar{Q}$ pair carries the transverse momentum, the predictive power of the yield is limited by the uPDFs. Therefore, to construct a meaningful uncertainty band, we fix the factorization scale at $\mu_F=m_T$. As we push toward the limit of the $k_T$-factorization approach with uPDFs at high $p_T$ at LO, we can only improve the high $p_T$ limit by a full NLO calculation.

After fixing the factorization scale, the variation in renormalization scale then gives the largest uncertainty, followed by the variation in charm mass. When $\mu_R$ is reduced, the strong coupling constant is larger, increasing the yield. On the other hand, when $m_c$ is reduced, the yield increases. In the remainder of this section, we present our results by adding the uncertainties due to variations of the charm mass and renormalization scale in quadrature. 

\begin{figure*}
\centering
\begin{minipage}[ht]{0.97\columnwidth}
\centering
\includegraphics[width=\columnwidth]{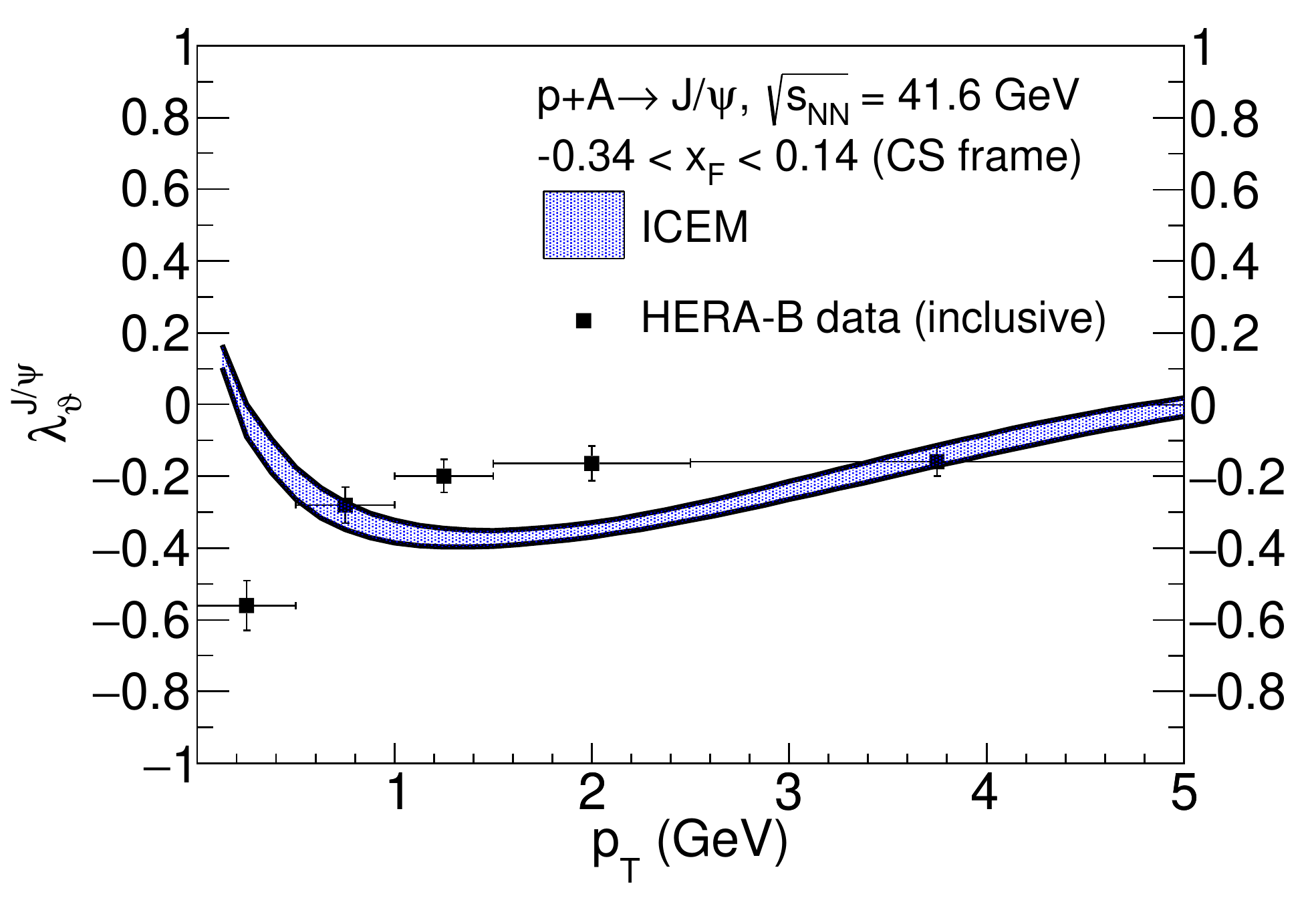}
\caption{The $p_T$ dependence of the polarization parameter $\lambda_\vartheta$ for prompt $J/\psi$ production in the Collins-Soper frame at $\sqrt{s_{NN}} = 41.6$~GeV in the ICEM with mass uncertainties are compared to the HERA-B data for inclusive $J/\psi$ \cite{Abt:2009nu}.} \label{HERAB_CS}
\end{minipage}%
\hspace{1cm}
\begin{minipage}[ht]{0.97\columnwidth}
\centering
\includegraphics[width=\columnwidth]{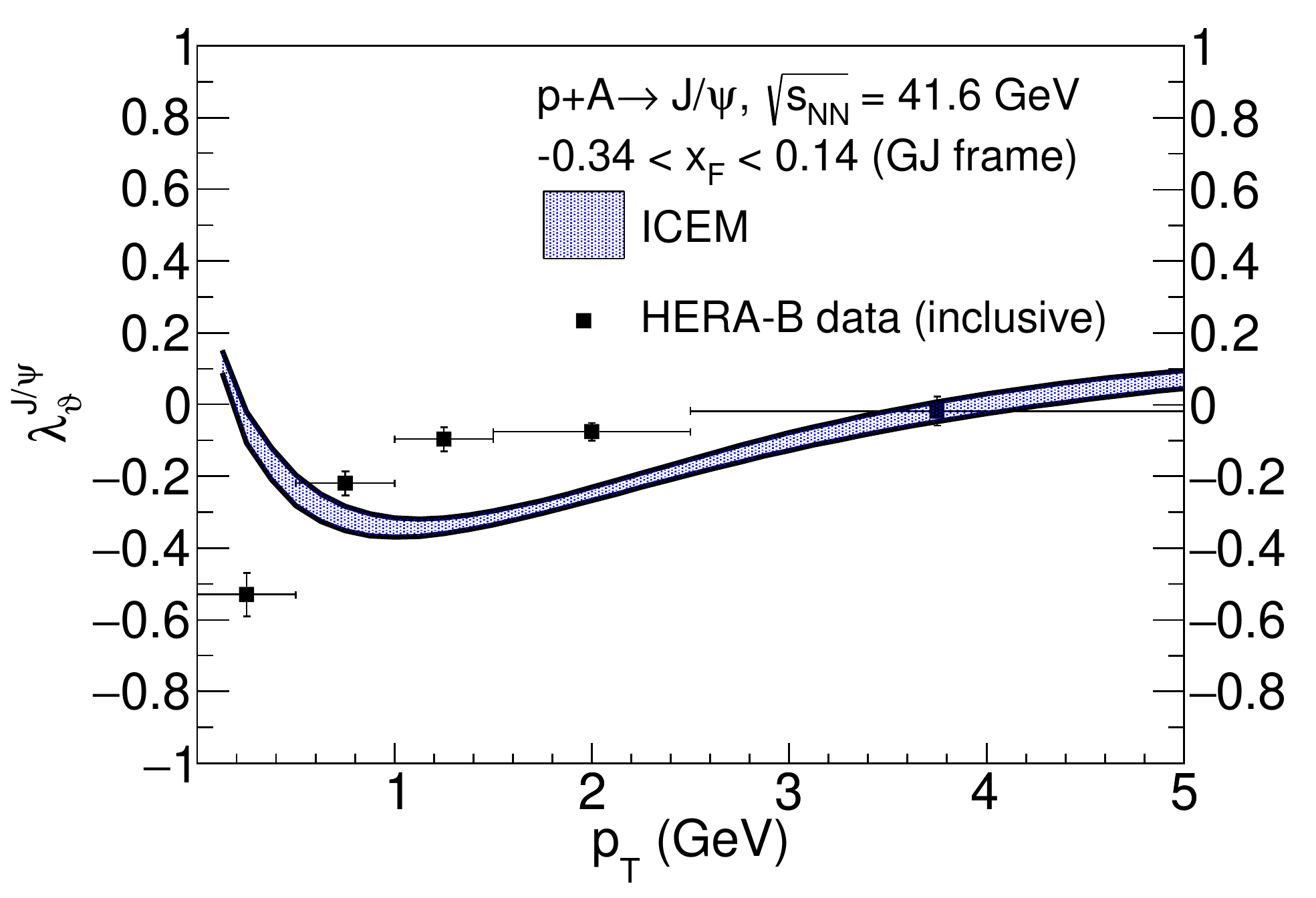}
\caption{The $p_T$ dependence of the polarization parameter $\lambda_\vartheta$ for prompt $J/\psi$ production in the Gottfried-Jackson frame at $\sqrt{s_{NN}} = 41.6$~GeV in the ICEM with mass uncertainties are compared to the HERA-B data for inclusive $J/\psi$ \cite{Abt:2009nu}.} \label{HERAB_GJ}
\end{minipage}
\end{figure*}

The inclusive $J/\psi$ $p_T$ distribution at $\sqrt{s} = 7$~TeV with combined uncertainty is shown in Fig.~\ref{LHCb_1S_pt}. The ICEM result has a peak at $p_T$$\sim$2~GeV, in agreement with the experimental results but slightly overestimates the data at high $p_T$. The ICEM $p_T$ distribution is within reasonable agreement with the data for all $p_T$. The experimental prompt production cross section depends on the polarization of $J/\psi$ since the polarization affects the acceptance and reconstruction efficiencies. LHCb checked the yields for the three polarization assumptions: $\lambda_\vartheta = -1,0,+1$. The experimental $p_T$ distribution for all polarization assumptions is within the uncertainty band constructed in the ICEM. By matching to the experimental unpolarized yield $\lambda_\vartheta=0$, we find that the ICEM can describe the $J/\psi$ $p_T$ distribution with $F_{J/\psi}=0.0216$. This is the fraction of $c\bar{c}$ pairs produced in the invariant mass range from $M_{J/\psi}$ to $2m_{D^0}$ that result in direct $J/\psi$, defined in Eq.~(\ref{cem_sigma}).

We test the universality of $F_{J/\psi}$ by comparing the inclusive $J/\psi$ $p_T$ distribution in the ICEM at $\sqrt{s}=1.96$~TeV and $|y|<0.6$ with the CDF data \cite{Acosta:2004yw} in Fig.~\ref{LHCb_1S_pt}. We again assume the direct production takes a constant fraction of $0.62$ of the inclusive production \cite{Digal:2001ue} to obtain the inclusive $J/\psi$ cross section. The ICEM results slightly overshoot the data at high $p_T$ because both the direct and non-prompt contributions to $J/\psi$ production are $p_T$ dependent \cite{Andronic:2015wma,Aaij:2011jh}. The direct-to-prompt $J/\psi$ ratio decreases as $p_T$ grows and the contribution from $b$ decay to inclusive production is measured to be larger at high $p_T$ than at low $p_T$. Combining the effects of both, using a constant direct-to-inclusive ratio of $0.62$ gives an overestimate of the yields at high $p_T$. The calculated cross section differs from the measurements more as $p_T$ increases. We note that if we fix $F_{J/\psi}$ from the CDF data alone, it agrees within 1.5\% of that extracted from comparison to the LHCb data.

\subsubsection{$\psi$(2S) $p_T$ distribution}

The inclusive $\psi$(2S) $p_T$ distribution at $\sqrt{s} = 1.96$~TeV is shown in Fig.~\ref{CDF_2S_pt}. Here, the direct production cross section is calculated using Eq.~(\ref{pt_rap_cut}) by integrating the pair invariant mass from $M_{\psi{\rm (2S)}}$ to $2m_{D^0}$ over the rapidity range $|y|<0.6$. We assume the direct production is the same as the prompt production as there are no quarkonium states that feed down to $\psi$(2S) since its mass is just below $2m_{D^0}$. Therefore, we compare the $p_T$-integrated yield of direct $\psi$(2S) with the CDF measurement \cite{Aaltonen:2009dm}. We find $F_{\psi{\rm (2S)}}=0.117$. We note that $F_{\psi{\rm (2S)}} > F_{J/\psi}$, primarily because the mass range is much smaller for $\psi$(2S) than $J/\psi$. In the traditional CEM, $F_{\psi{\rm (2S)}}$ is smaller than $F_{J/\psi}$ because the integration over the pair invariant mass is the same for both $J/\psi$ and $\psi$(2S). We add the contribution from non-prompt production reported by the CDF Collaboration to our prompt production yield to give the inclusive $\psi$(2S) yield shown in Fig.~\ref{CDF_2S_pt}. We find agreement with the data within the combined uncertainty band constructed by varying the charm mass and the renormalization scale in the ICEM.

\subsubsection{$\chi_{\rm c1}$ and $\chi_{\rm c2}$ $p_T$ distribution}

We now turn to the $p_T$ dependence of $\chi_c$ production. The $p_T$ distributions of direct $\chi_{c1}$, direct $\chi_{c2}$, and the ratio of $\chi_{c2}$ to $\chi_{c1}$ at $\sqrt{s}=7$~TeV and 13~TeV are presented in Fig.~\ref{ATLAS_chic}. The direct production is calculated using Eq.~(\ref{pt_rap_cut}) by integrating the pair invariant mass from $M_{\chi_c}$ to $2m_{D^0}$ ($m_{D^0}=1.86$~GeV) over the rapidity range $|y|<0.75$. We assume the prompt production of $\chi_{c}$ is approximately the same as the direct production. Thus, by comparing the direct $\chi_{c1}$ and $\chi_{c2}$ yields in the ICEM with the experimental yield of prompt $\chi_{c1}$ and $\chi_{c2}$ at $\sqrt{s}=7$~TeV measured by the ATLAS Collaboration \cite{ATLAS:2014ala}, we obtain $F_{\chi_{c1}}=0.180$ and $F_{\chi_{c2}}=0.20$. As is the case for $F_{\psi{\rm (2S)}}$ and $F_{J/\psi}$, $F_{\chi_{c2}} > F_{\chi_{c1}}$ is because the integration range over the pair invariant mass is smaller for $\chi_{c2}$ than for $\chi_{c1}$. In the tradition CEM, $F_{\chi_{c2}}$ is smaller than $F_{\chi_{c1}}$. The direct production in the ICEM describes prompt production of both $\chi_{c1}$ and $\chi_{c2}$ at $\sqrt{s}=7$~TeV within the uncertainty bands constructed by varying the charm quark mass and renormalization scale. The ratio of the cross sections is also described by the ICEM. We calculate the $\chi_{c2}$ to $\chi_{c1}$ ratio to be $\sim 0.5$, almost independent of $p_T$. The ratios disagree with a recent NRQCD calculation \cite{Cisek:2017gno}, which the ratio decreases as $p_T$ increases and is above the data. We assume that $p_{T\chi_c}\approx p_{TJ/\psi}$, not unreasonable since the mass difference is $\sim$500~MeV and the decay photon is soft. We anticipate the direct $\chi_{c1}$ and $\chi_{c2}$ yields will be increased by 51\% (at $p_T=10$~GeV) to 120\% (at $p_T=30$~GeV) when $\sqrt{s}$ is increased from 7~TeV to 13~TeV. However, the ratio of $\chi_{c2}$ to $\chi_{c1}$ should remain approximately the same.

\begin{figure*}
\centering
\begin{minipage}[ht]{0.97\columnwidth}
\centering
\includegraphics[width=\columnwidth]{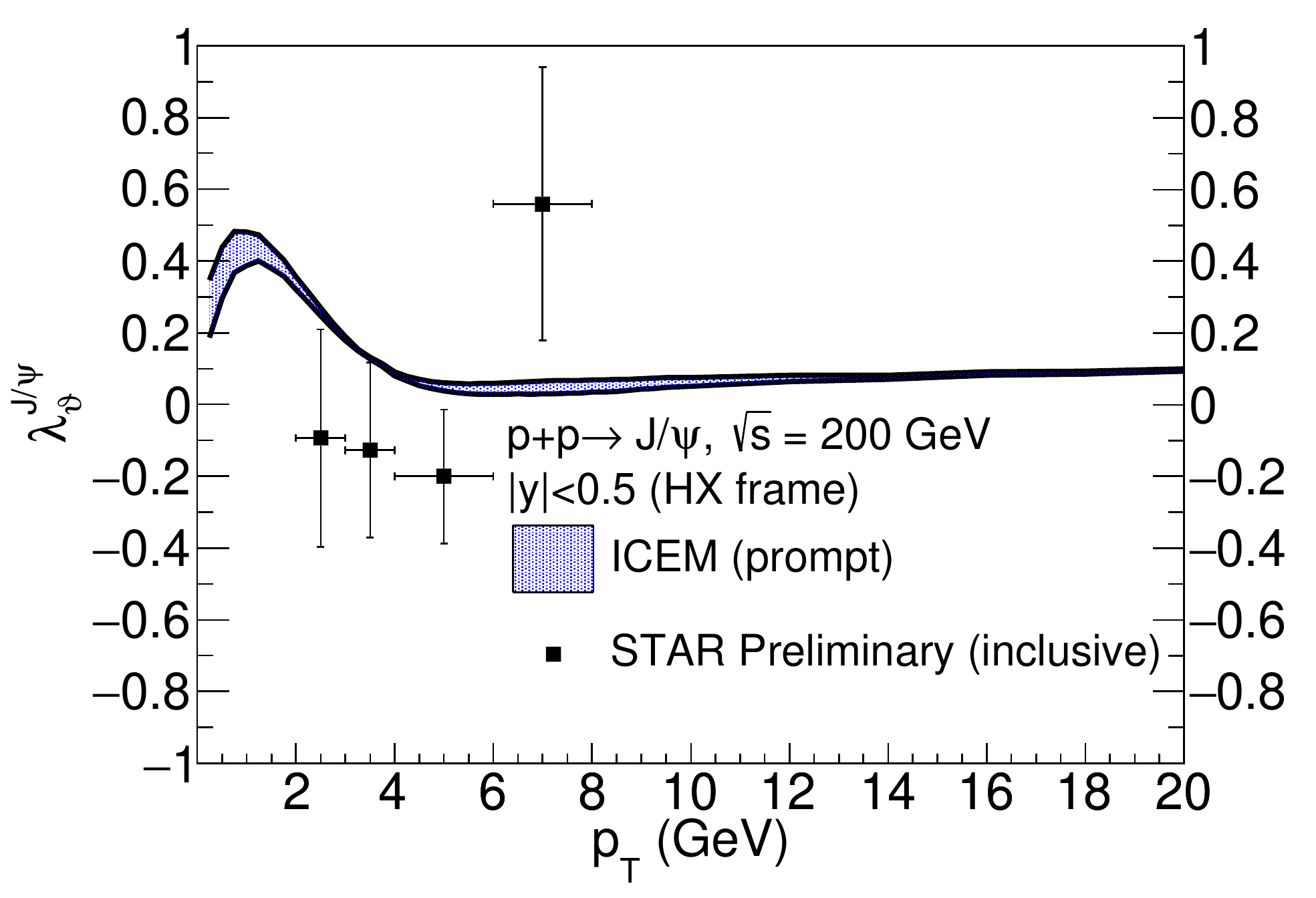}
\caption{The $p_T$ dependence of the polarization parameter $\lambda_\vartheta$ for prompt $J/\psi$ production at $\sqrt{s} = 200$~GeV in the ICEM with mass uncertainty. The STAR data for inclusive $J/\psi$ are also shown.} \label{STAR_HX}
\end{minipage}%
\hspace{1cm}%
\begin{minipage}[ht]{0.97\columnwidth}
\centering
\includegraphics[width=\columnwidth]{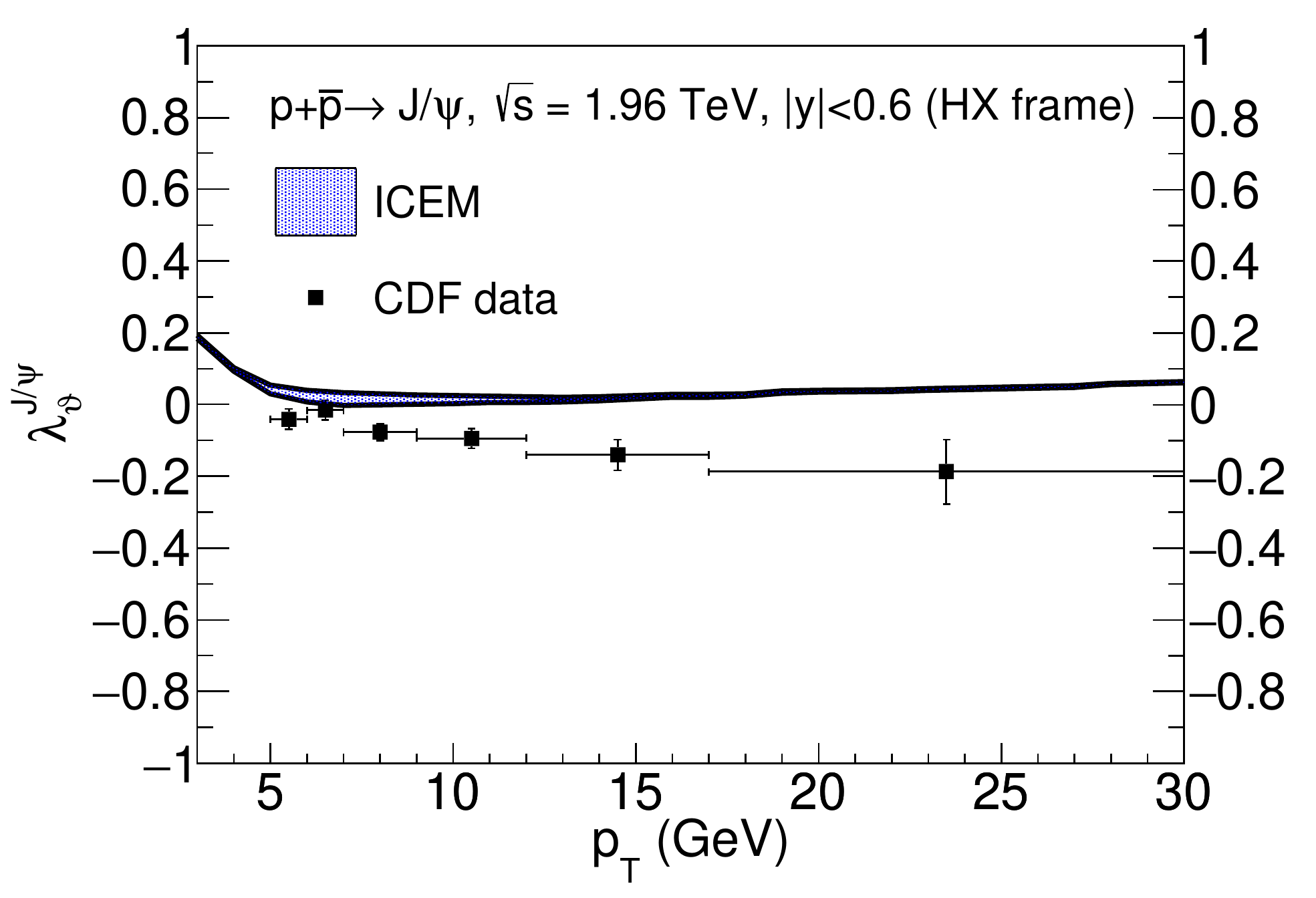}
\caption{The $p_T$ dependence of the polarization parameter $\lambda_\vartheta$ for prompt $J/\psi$ production at $\sqrt{s} = 1.96$~TeV in the ICEM with mass uncertainty. The CDF data are also shown \cite{Abulencia:2007us}.} \label{CDF_HX}
\end{minipage}
\end{figure*}

\subsubsection{Prompt $J/\psi$ $p_T$ distribution}
After fixing $F_{J/\psi}$, $F_{\psi{\rm (2S)}}$, $F_{\chi_{c1}}$ and $F_{\chi{c2}}$, we calculate the prompt $J/\psi$ $p_T$ distribution at $\sqrt{s}=7$~TeV in the rapidity range $2.0 < y < 4.5$ using the direct $J/\psi$, $\psi$(2S), $\chi_{c1}$ and $\chi_{c2}$ yields and their branching ratios to $J/\psi$. The prompt $J/\psi$ $p_T$ distribution is shown in Fig.~\ref{LHCb_1S_prompt}. The ICEM $p_T$ distribution describes the data for most $p_T$ but overshoots the data slightly at the highest $p_T$ bin. The ICEM $p_T$ distribution is within reasonable agreement with the data for all $p_T$. We extract the $p_T$ dependent feed-down ratios $c_{\psi}$'s by taking the direct to prompt ratio in this distribution. We find the feed-down ratios are very similar to those listed in Table~\ref{states}. Additionally, we find $c_{J/\psi}$ decreases as $p_T$ increases, in agreement with Ref.~\cite{Andronic:2015wma}.

\subsubsection{$J/\psi$ rapidity distribution}

We now turn to the rapidity dependence of $J/\psi$ production. The rapidity distribution of inclusive of $J/\psi$ at $\sqrt{s} = 7$~TeV is shown  in Fig.~\ref{ALICE_1S_rapidity}. The direct production is calculated using Eq.~(\ref{y_pt_cut}) by integrating over the $p_T$ range $0<p_T<7$~GeV ($|y|<0.9$) and $0<p_T<8$~GeV ($2.5<y<4$). We again assume the direct production is a constant 62\% \cite{Digal:2001ue} of the inclusive production. We use the same $F_{J/\psi}$ again to compare the rapidity distribution in the ICEM with the measurement made by the ALICE Collaboration \cite{Aamodt:2011gj}. The difference in the integrated $p_T$ range has a negligible on the rapidity distribution because the $p_T$ dependence has already dropped by an order of magnitude by $p_T\sim7-8$~GeV. We find the ICEM can describe the ALICE rapidity distribution at $\sqrt{s}=7$~TeV using the $F_{J/\psi}$ obtained at the same energy by LHCb in the forward rapidity region.
\subsubsection{$\psi$(2S) rapidity distribution}

The rapidity distribution of direct $\psi$(2S) at $\sqrt{s}=7$~TeV is shown in Fig.~(\ref{ALICE_2S_rapidity}). Here, the rapidity distribution is calculated in the interval $p_T<12$~GeV at forward rapidity ($2.5 < y < 4$).  We use the same $F_{\psi{\rm (2S)}}$ compare with inclusive $\psi$(2S) data from ALICE \cite{Abelev:2014qha}. While the lower bound of our uncertainty band should still be lower than the data when the contribution from $B$ decays are added, our baseline should slightly overshoot the inclusive $\psi$(2S) data. Our results also agree with the direct $\psi$(2S) rapidity distribution from a recent NRQCD calculation at LO using the $k_T$-factorization approach \cite{Cisek:2017gno}.

\subsection{$p_T$ dependence of $\lambda_\vartheta$}

Here, we present the $p_T$ dependence of the polarization parameter $\lambda_\vartheta$ in $p+p$ and $p$+A collisions. Because the polarization parameter is defined as the ratio of polarized to unpolarized cross sections in Eq.~(\ref{mix_psi}) and these cross sections depend on $\mu_R$ and $\mu_F$ in the same way, the polarization parameter is independent of the scale choice. However, the amplitudes themselves are mass dependent so that the polarized to unpolarized ratio in $\lambda_\vartheta$ depends on the charm quark mass. Thus the only uncertainty on $\lambda_\vartheta$ in our calculation is due to the variation of $m_c$ in the range $1.2 < m_c < 1.5$~GeV. In this section, the uncertainty band is only due to the mass variation and therefore the uncertainty is reduced relative to the yield calculations.

\begin{figure*}
\centering
\begin{minipage}[ht]{0.97\columnwidth}
\centering
\includegraphics[width=\columnwidth]{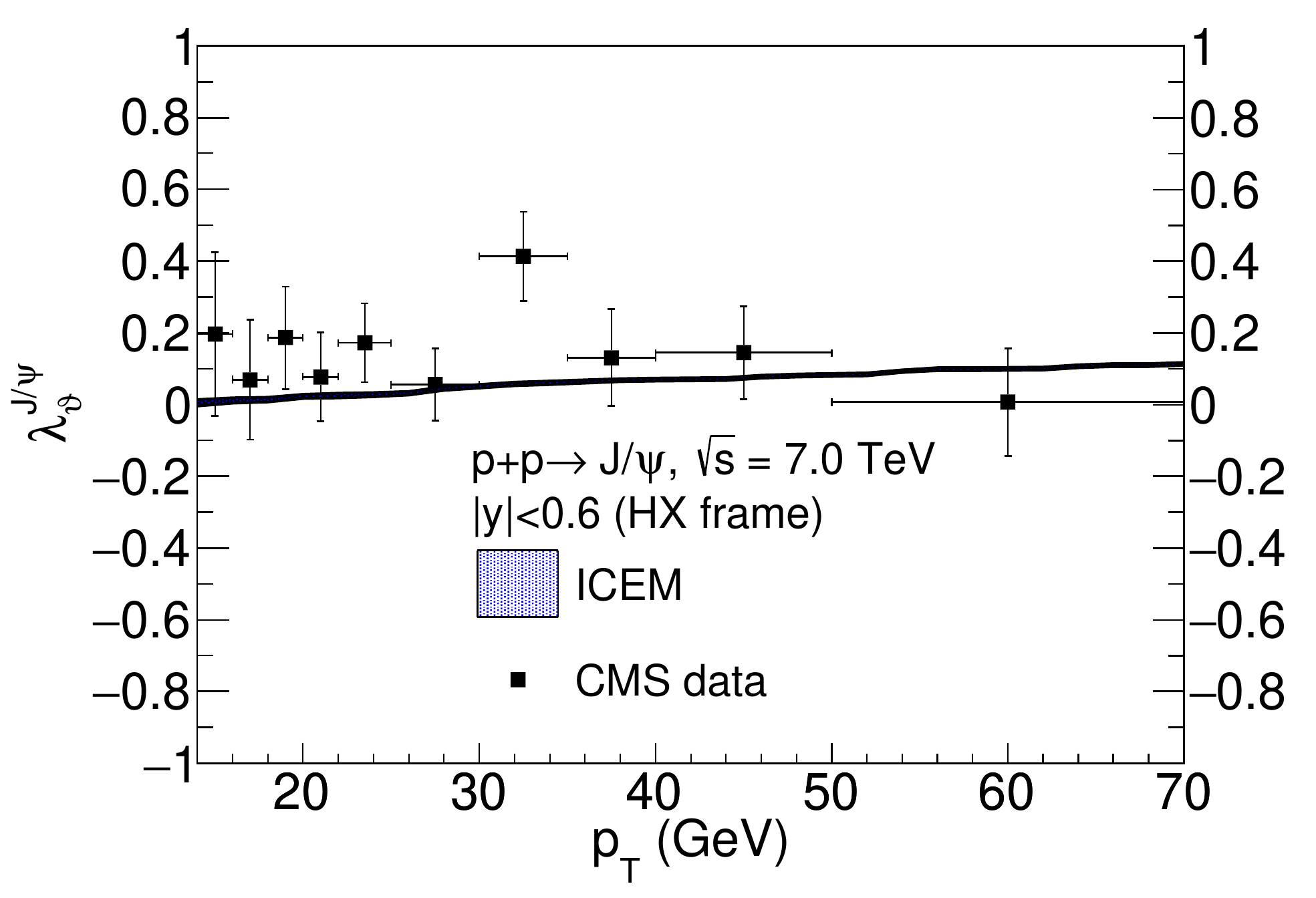}
\caption{The $p_T$ dependence of the polarization parameter $\lambda_\vartheta$ for prompt $J/\psi$ production at $\sqrt{s} = 7$~TeV in the region $|y|<0.6$ in the ICEM with mass uncertainty. The CMS data are also shown  \cite{Chatrchyan:2013cla}.} \label{CMS_HX_central}
\end{minipage}%
\hspace{1cm}%
\begin{minipage}[ht]{0.97\columnwidth}
\centering
\includegraphics[width=\columnwidth]{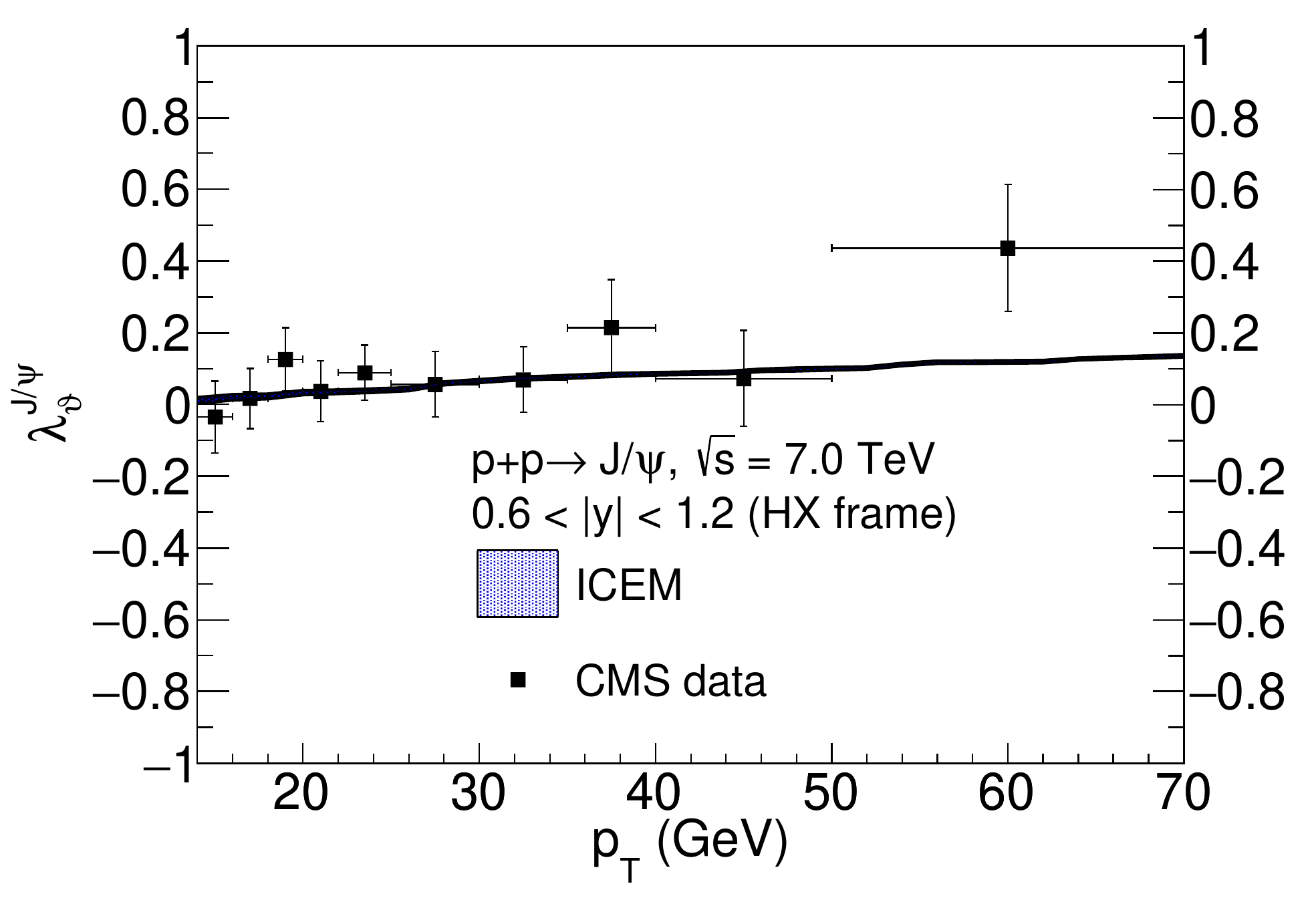}
\caption{The $p_T$ dependence of the polarization parameter $\lambda_\vartheta$ for prompt $J/\psi$ production at $\sqrt{s} = 7$~TeV in the region $0.6<|y|<1.2$ in the ICEM with mass uncertainty. The CMS data are also shown \cite{Chatrchyan:2013cla}.} \label{CMS_HX_forward}
\end{minipage}
\end{figure*}

\begin{figure}[hb]
\centering
\includegraphics[width=\columnwidth]{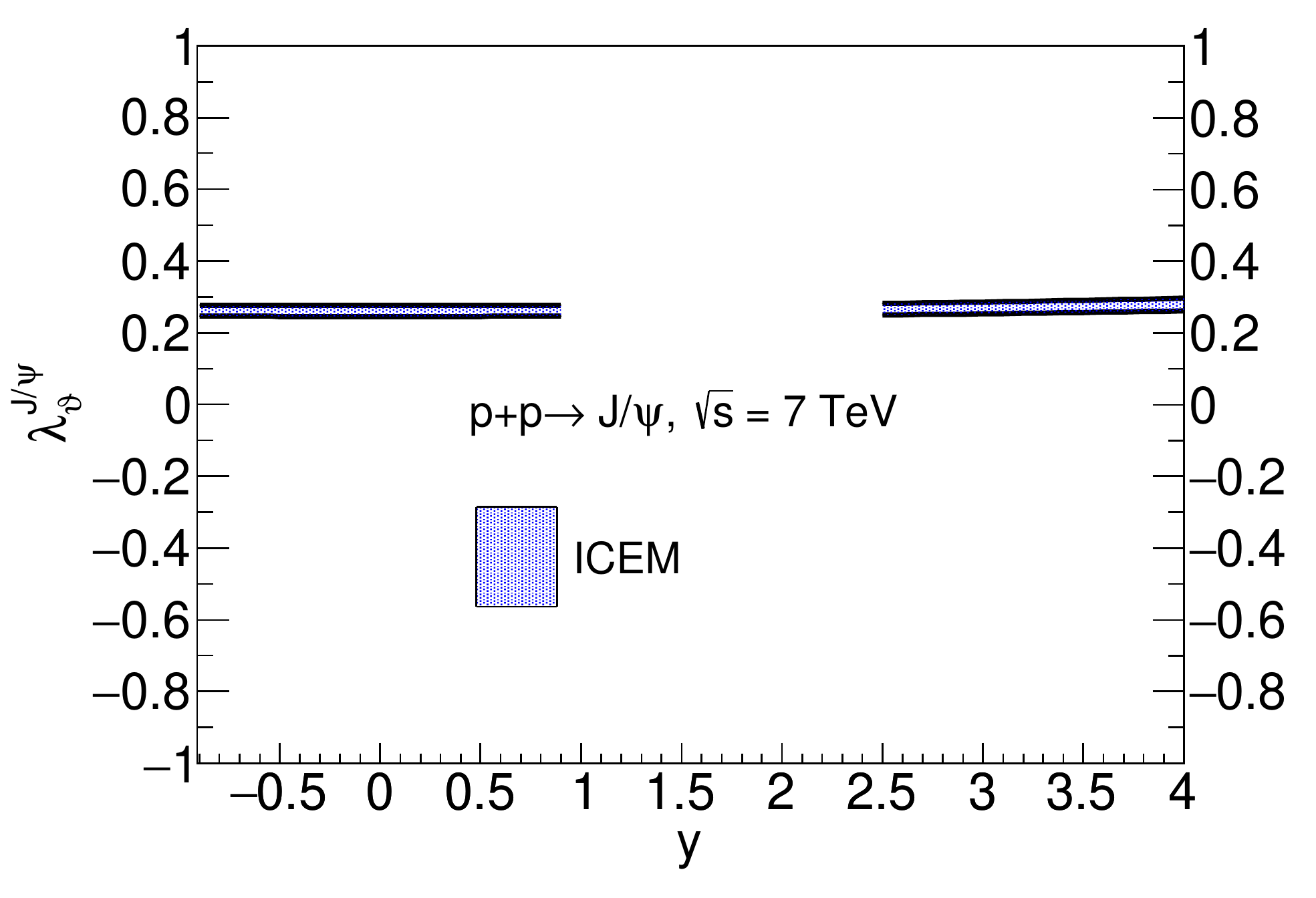}
\caption{The $p_T$ integrated rapidity dependence of $\lambda_\vartheta$ for prompt $J/\psi$ production at $\sqrt{s} = 7$~TeV in the helicity frame in the ALICE acceptance. Note that we use the same kinematic cuts as on the yields in Fig.~\ref{ALICE_1S_rapidity}.} \label{ALICE_y}
\end{figure}

\subsubsection{Charmonium polarization in $p$+A collisions at fixed-target energies}

The polarization results for prompt production of $J/\psi$ at $\sqrt{s_{NN}} = 41.6$~GeV are shown in Figs.~\ref{HERAB_CS} and \ref{HERAB_GJ}. Although the HERA-B data are taken on nuclear targets, C and W, and there are known nuclear modifications of the parton densities in the nucleus, $\lambda_\vartheta$ is independent of any modification. This is because the ratios of the polarized to unpolarized cross sections are in the same kinematic acceptance and any nuclear effects cancel in the ratio. Thus there is no difference in polarization between the two target nuclei. We compare our results with the C and W combined data measured by the HERA-B Collaboration in the region $-0.34 < x_F < 0.14$ \cite{Abt:2009nu}.

Prompt $J/\psi$ polarization in the ICEM is close to unpolarized in both the CS and GJ frames for $p_T < 5$~GeV. At $p_T = 0$, the two $z$-axes $z_{\rm CS}$ and $z_{\rm GJ}$, are in the same direction. Thus the polarization is the same in that limit. As $p_T$ increases, the two axes depart from each other. Thus the polarization is slightly less longitudinal in the GJ frame than in the CS frame. This behavior is also consistent with the experimental data showing that the $J/\psi$ polarization at very low $p_T$ is not affected by switching from the CS frame to the GJ frame. At higher $p_T$ the polarization is slightly less longitudinal in the GJ frame than in the CS frame. The ICEM results are in fair agreement with the experimental data except at the lowest $p_T$. 

\subsubsection{Charmonium polarization in $p$+$p$($\bar{p}$) collisions}

We present the polarization parameters for prompt $J/\psi$ in $p$+$p$ collisions at $\sqrt{s}=200$~GeV in Fig.~\ref{STAR_HX}. We compare our results with the data from the STAR Collaboration in the region $|y|<0.5$ \cite{STAR.MA} in the helicity frame. The ICEM polarization of prompt $J/\psi$ in the helicity frame is slightly transverse at low $p_T$ ($p_T<M_{J/\psi}$). The result becomes unpolarized at moderate $p_T$ ($M_{J/\psi} < p_T < 2M_{J/\psi}$) before changing to slightly transverse at high $p_T$. The ICEM polarization agrees fairly well with the data at small and moderate $p_T$ for inclusive $J/\psi$ polarization at STAR.

We also compared the polarization parameters for prompt $J/\psi$ in $p$+$\bar{p}$ collisions at $\sqrt{s}=1.96$~TeV with the data measured by the CDF Collaboration in the region $|y|<0.6$ \cite{Abulencia:2007us} in the helicity frame, shown in Fig~\ref{CDF_HX}. The ICEM prompt $J/\psi$ polarization does not depend strongly on $\sqrt{s}$ or whether the collision is $p$+$p$ or $p$+$\bar{p}$. We find the trend in the $p_T$ dependence of the polarization is the same.  At high $p_T$, the prompt $J/\psi$ polarization measured by the CDF Collaboration is slightly longitudinal to unpolarized while the ICEM polarization is slightly transverse. The polarization predicted by NRQCD also shows a similar behavior at this energy \cite{Braaten:1999qk}. However, NRQCD predicts a stronger transverse polarization ($\lambda_\vartheta$$\sim$0.6) than ICEM in the high $p_T$ limit.


\subsection{Rapidity dependence of $\lambda_\vartheta$}

Next we turn to the rapidity dependence of $\lambda_\vartheta$. We calculate the prompt $J/\psi$ polarization in the helicity frame for $p+p$ collisions at $\sqrt{s}=7$~TeV in two rapidity ranges, $|y|<0.6$ and $0.6<|y|<1.2$, shown in Figs.~\ref{CMS_HX_central} and \ref{CMS_HX_forward} respectively. We compare our results to the experimental data from the CMS Collaboration \cite{Chatrchyan:2013cla}. There is no difference in the polarization of prompt $J/\psi$ in these two rapidity regions in the ICEM. In the ICEM, the polarization parameter $\lambda_\vartheta$ of prompt $J/\psi$ production increases very slowly in the high $p_T$ limit and reaches $\lambda_\vartheta \sim 0.12$ at $p_T=70$~GeV. The ICEM polarization agrees with the the experimental results at central rapidity within uncertainty except the data in the $30<p_T<35$ bin. The experiment reports the polarization is less transverse in the forward rapidity region. Our results in the ICEM still agrees with the data even though the calculated polarization does not depend on rapidity in this range at 7~TeV.

\begin{figure*}
\centering
\begin{minipage}[ht]{0.97\columnwidth}
\centering
\includegraphics[width=\columnwidth]{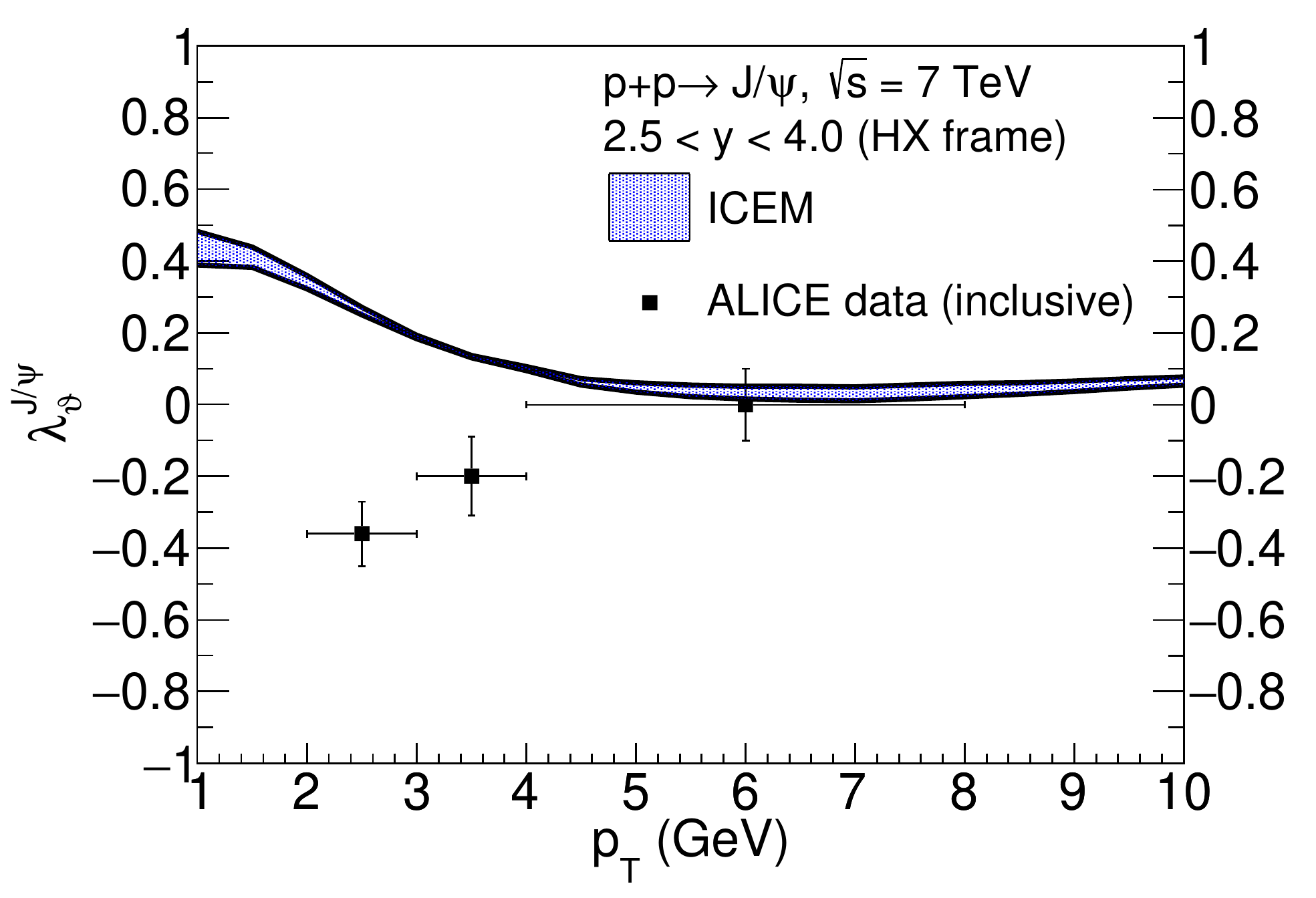}
\caption{The $p_T$ dependence of $\lambda_\vartheta$ for prompt $J/\psi$ production at $\sqrt{s} = 7$~TeV in the helicity frame is compared with the ALICE data for inclusive $J/\psi$ \cite{Abelev:2011md}.} \label{ALICE_HX}
\end{minipage}%
\hspace{1cm}%
\begin{minipage}[ht]{0.97\columnwidth}
\centering
\includegraphics[width=\columnwidth]{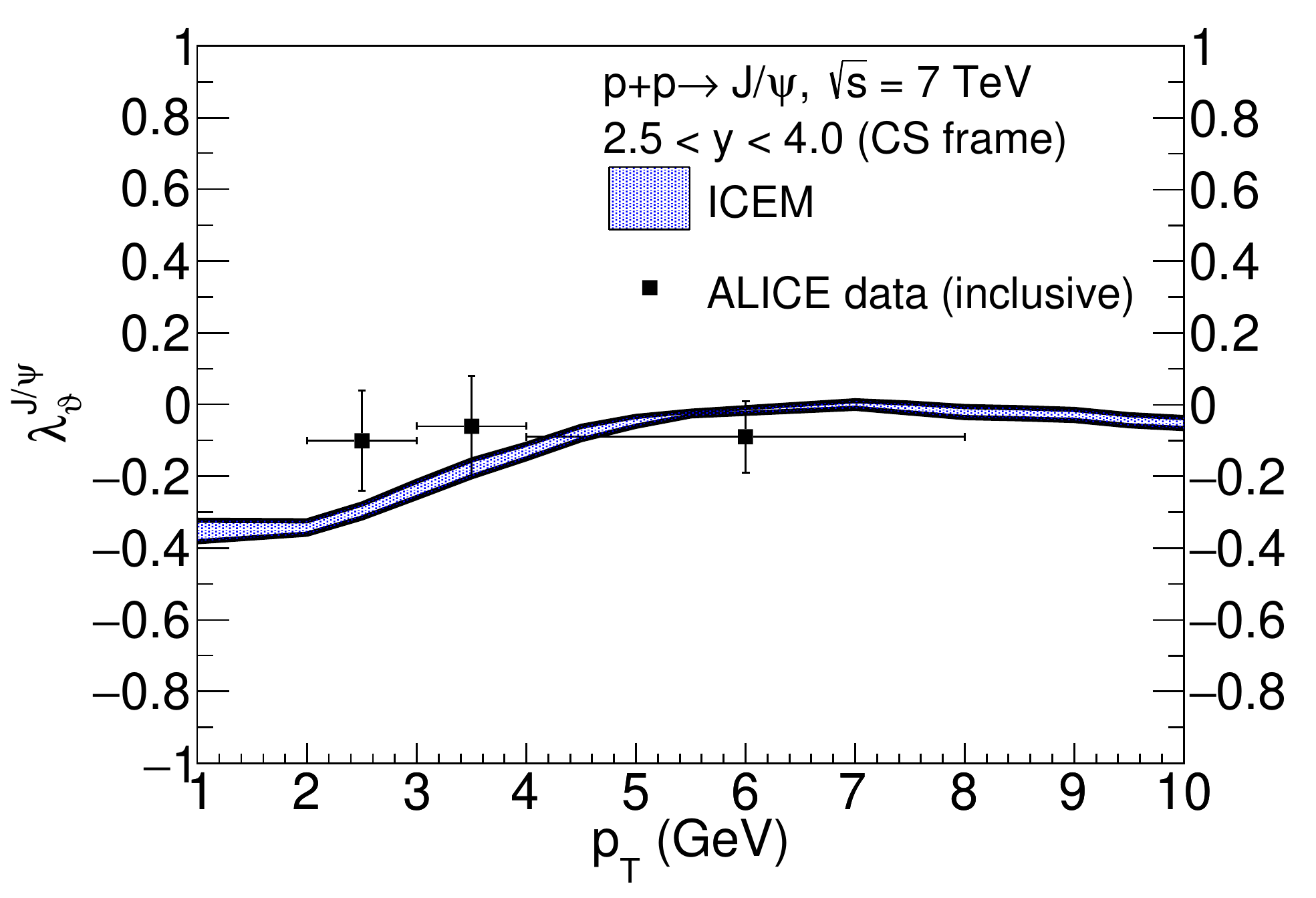}
\caption{The $p_T$ dependence of $\lambda_\vartheta$ for prompt $J/\psi$ production at $\sqrt{s} = 7$~TeV in the Collins-Soper frame is compared with the ALICE data for inclusive $J/\psi$ \cite{Abelev:2011md}.} \label{ALICE_CS}
\end{minipage}
\end{figure*}

We also do not observe variations in the polarization parameter $\lambda_\vartheta$ at $\sqrt{s}=7$~TeV in the region of $y<4$ using the same kinematics cut compared to the ALICE yield measurement in Fig.~\ref{ALICE_1S_rapidity}. We present the polarization as a function of rapidity in Fig.~\ref{ALICE_y}. The polarization parameter of prompt $J/\psi$ for the $p_T$-integrated results is $\lambda_\vartheta = 0.26\pm0.02$.


\subsection{Frame dependence of $\lambda_\vartheta$}

We now turn to the frame dependence of our 7~TeV results. We calculate the polarization parameter in $p+p$ collisions at $\sqrt{s}=7$~TeV in both the helicity frame and the Collins-Soper frame, shown in Figs.~\ref{ALICE_HX} and \ref{ALICE_CS} respectively. The polarization in the Collins-Soper frame is opposite to that in the helicity frame in the ICEM. We expect this because, in these kinematics, at order $\alpha_s^2$, the polarization axis in the Collins-Soper frame is always perpendicular to that in the helicity frame. Therefore, at low $p_T$, where the $J/\psi$ is predicted to be slightly transverse in the helicity frame, it is predicted to be slightly longitudinal in the Collins-Soper frame. Whereas, at moderate $p_T$, where the $J/\psi$ is predicted to be unpolarized, it is also predicted to be unpolarized in the Collins-Soper frame. This behavior, however, is not measured experimentally. As we compare our results with the ALICE data \cite{Abelev:2011md}, the ICEM polarization agrees with the data in the Collins-Soper frame but does not agree with the data in the helicity frame, especially at low $p_T$ where the frame dependence is most significant.

We find similar results by comparing to the LHCb data in the Collins-Soper frame \cite{Aaij:2013nlm}, show in in Figs.~\ref{LHCb_HX} and \ref{LHCb_CS}: the polarization in the ICEM agrees with the data in the Collins-Soper frame but not in the helicity frame. We expect that the difference in agreement of the calculations in different frames with the data may be resolved with a full $\alpha_s^3$ calculation of the ICEM cross section.

Finally, we note that at low $p_T$ the polarization in the Gottfried-Jackson frame is similar to that in the Collins-Soper frame, as shown in Figs.~\ref{HERAB_CS} and \ref{HERAB_GJ} for fixed-target energies. However at high $p_T$, the polarization in the Gottfried-Jackson frame is similar to that in the helicity frame. The differences are due to the definition of the polarization axes in the quarkonium rest frame. When $p_T<<m_T$, the angle between the polarization axis in the Gottfried-Jackson frame and that in the Collins-Soper frame is small. As $p_T$ increases, the polarization axis in the Gottfried-Jackson frame becomes collinear with that in the helicity frame. Therefore, the polarization calculated in the Gottfried-Jackson frame is opposite to that in the helicity frame at low $p_T$, and thus similar to that in the Collins-Soper frame. But as $p_T$ increases, the polarization in the Gottfried-Jackson frame should asymptotically approach the polarization in the helicity frame.


\begin{table}
\caption{\label{feed_down_sensitivity}Values of $c_\mathcal{Q}$ used to test the sensitivity of our results to the feed-down ratios. Based on the uncertainty in $c_\mathcal{Q}$ (third column), $c_\mathcal{Q}^\prime$ (second column) is used assuming the promptly produced 1S states comprise less directly produced 1S states, and $c_\mathcal{Q}^{\prime \prime}$ (fourth column) is used assuming the promptly produced 1S states comprise more directly produced 1S states,}
\begin{ruledtabular}
\begin{tabular}{cccc}
$\mathcal{Q}$ & $c_{\mathcal{Q}}^\prime$ & $c_{\mathcal{Q}}$ & $c_{\mathcal{Q}}^{\prime \prime}$\\
\hline
$J/\psi$ & 0.59 & 0.62$\pm$0.04 & 0.65  \\
$\psi$(2S) & 0.09 & 0.08$\pm$0.02 & 0.07 \\
$\chi_{c1}$(1P) & 0.17 & 0.16$\pm$0.04 & 0.15 \\
$\chi_{c2}$(1P) & 0.15 & 0.14$\pm$0.04 & 0.13 \\
\end{tabular}
\end{ruledtabular}
\end{table}

\begin{figure*}
\centering
\begin{minipage}[ht]{0.97\columnwidth}
\centering
\includegraphics[width=\columnwidth]{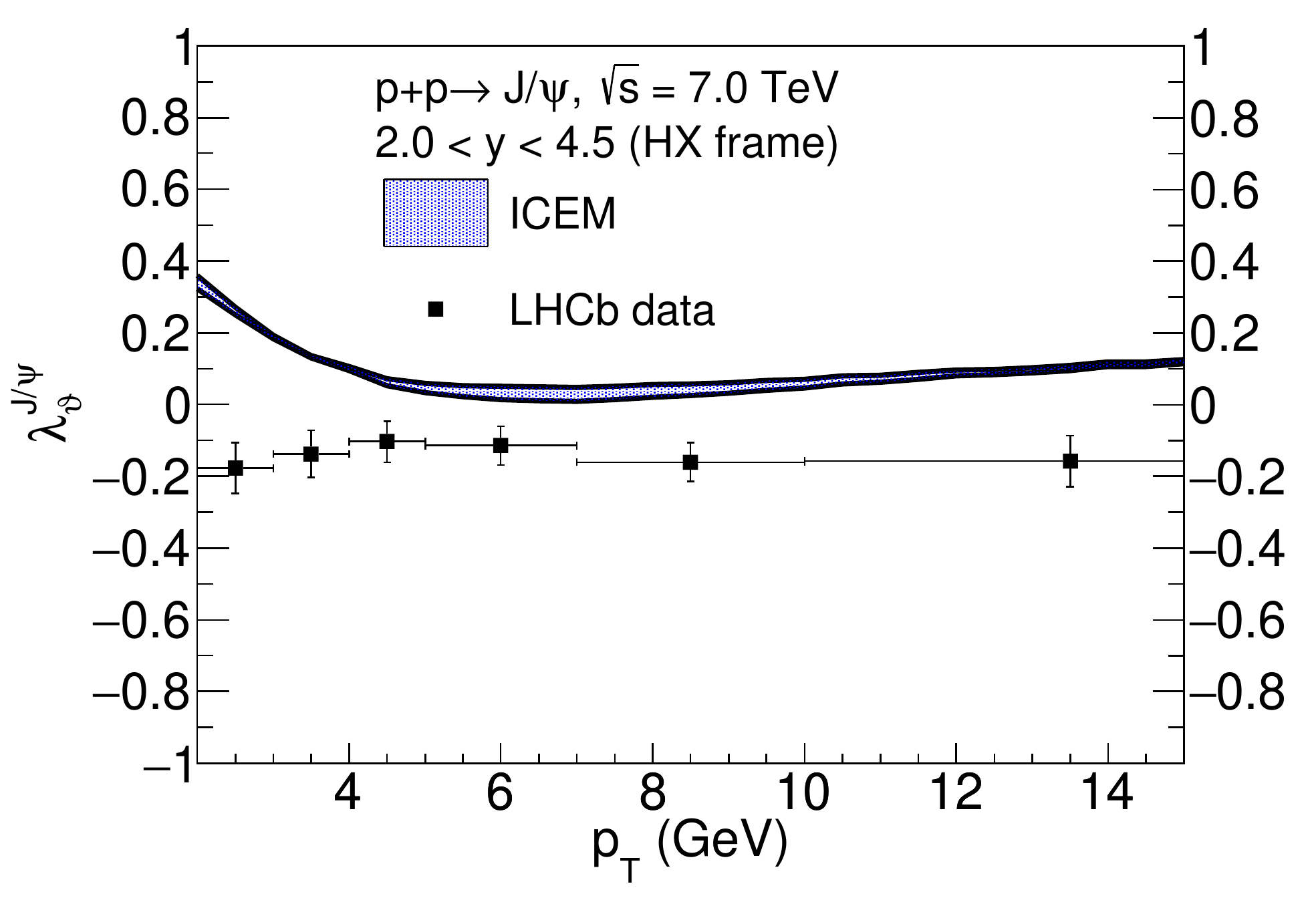}
\caption{The $p_T$ dependence of $\lambda_\vartheta$ for prompt $J/\psi$ production at $\sqrt{s} = 7$~TeV in the helicity frame is compared with the LHCb data \cite{Aaij:2013nlm}.} \label{LHCb_HX}
\end{minipage}%
\hspace{1cm}%
\begin{minipage}[ht]{0.97\columnwidth}
\centering
\includegraphics[width=\columnwidth]{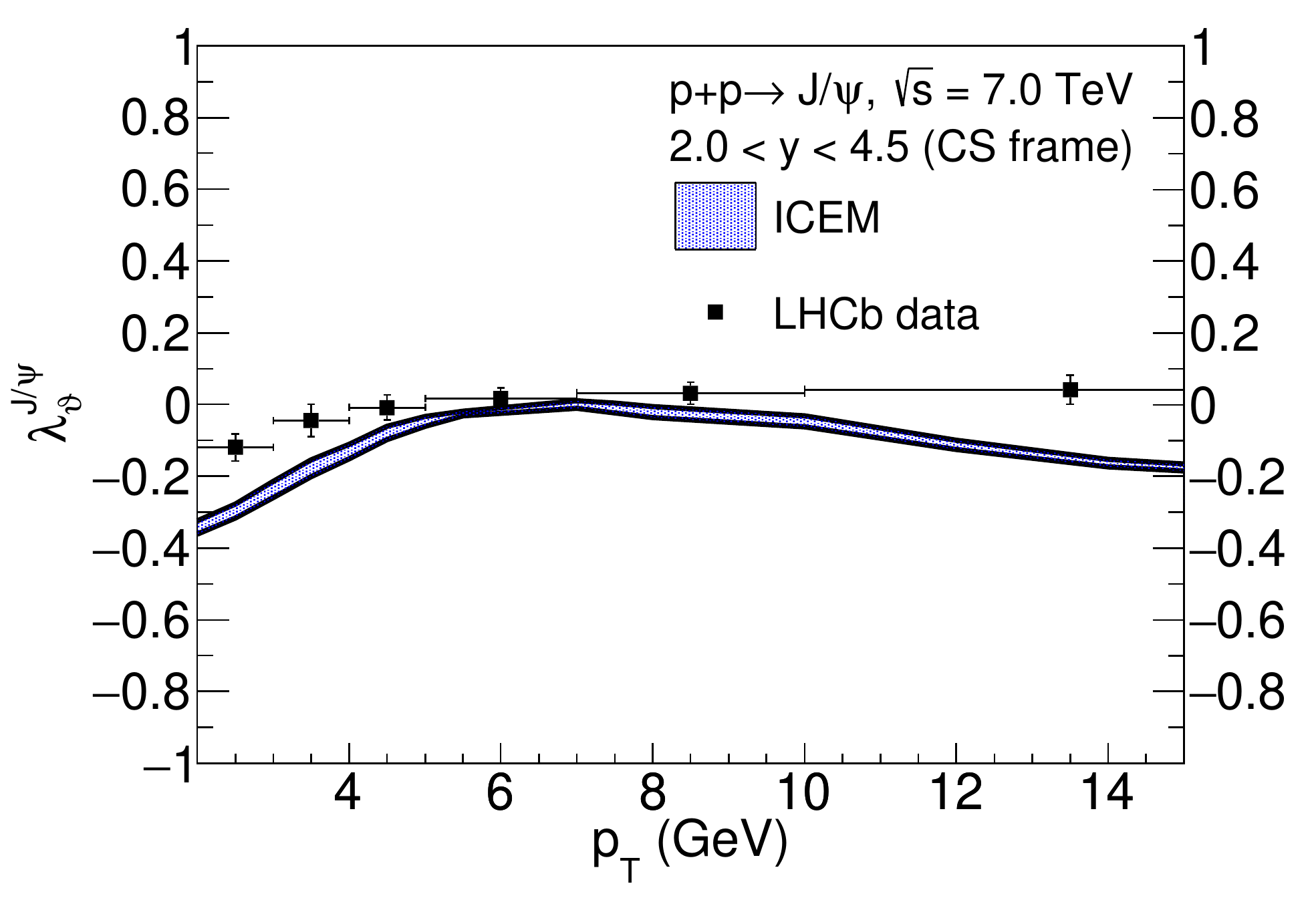}
\caption{The $p_T$ dependence of $\lambda_\vartheta$ for prompt $J/\psi$ production at $\sqrt{s} = 7$~TeV in the Collins-Soper frame is compared with the LHCb data \cite{Aaij:2013nlm}.} \label{LHCb_CS}
\end{minipage}
\end{figure*}

\begin{figure}[hb]
\centering
\includegraphics[width=\columnwidth]{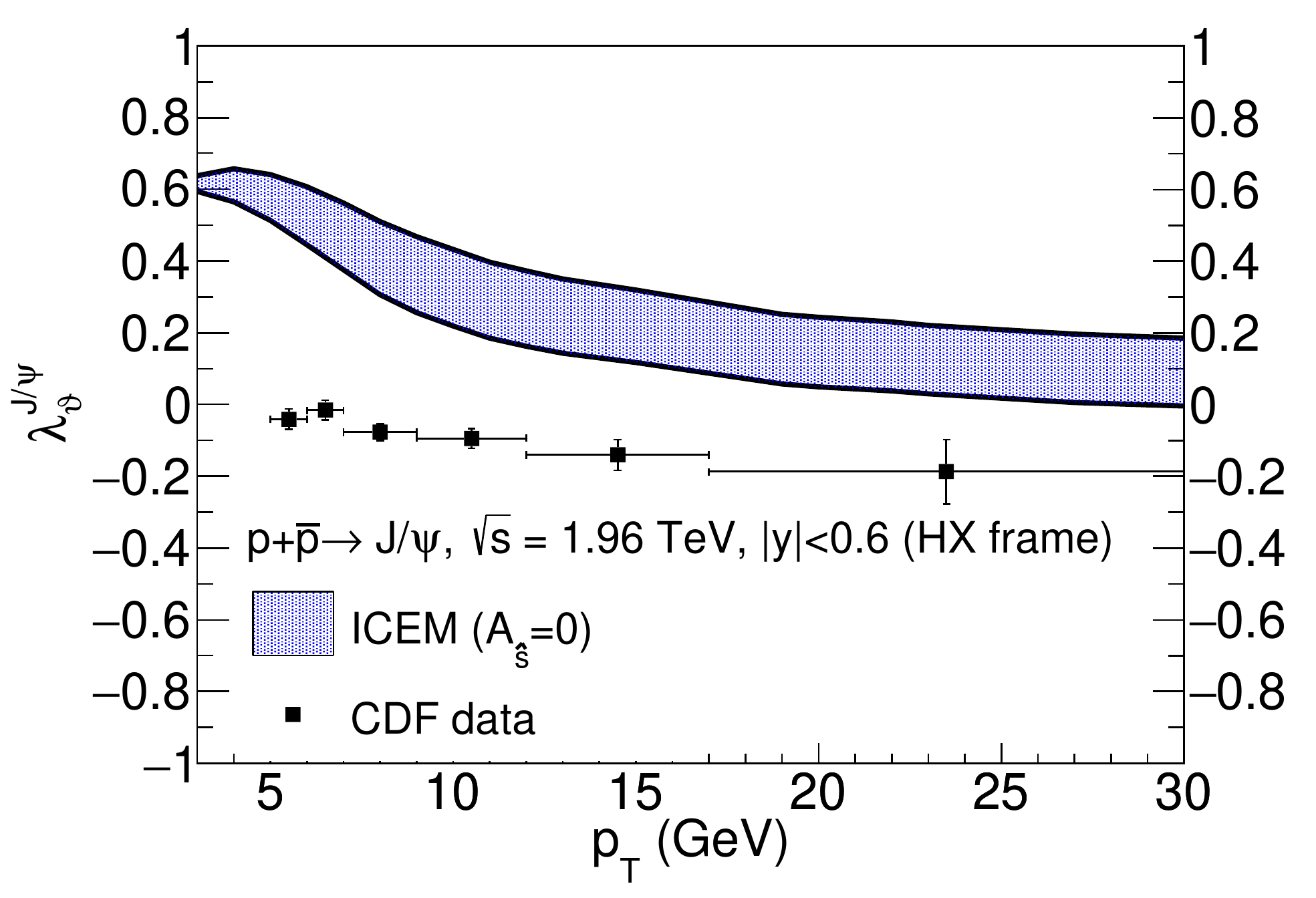}
\caption{The $p_T$ dependence of the polarization parameter $\lambda_\vartheta$ for prompt $J/\psi$ production at $\sqrt{s} = 1.96$~TeV in the ICEM with mass uncertainty when the $\hat{s}$-channel contribution is excluded. The CDF data are also shown \cite{Abulencia:2007us}.} \label{CDF_HX_tu}
\end{figure}

\subsection{Sensitivity to scales and quark mass}

We have already discussed the sensitivity of the charmonium yields to the factorization and the renormalization scales in section \ref{factorization}. Here we note that the longitudinal to unpolarized fraction $R^{J_z=0}_{J/\psi}$ used in the calculation of $\lambda_\vartheta$, is insensitive to scale variations because the longitudinal and transverse change similarly as the scales are varied. Therefore, the polarization parameter $\lambda_\vartheta$ for prompt $J/\psi$ is independent of the scales for all energies considered. Similarly, while the unpolarized $\chi_{c1}$ and $\chi_{c2}$ cross section vary appreciably with the scale choice, the $\chi_{c2}$ to $\chi_{c1}$ ratio is also independent of scales.

While the scale variations affect the polarized and unpolarized cross sections the same way, making $\lambda_\vartheta$ scale independent, the $J_z$ components of the polarized cross section depend differently on quark mass. When $p_T \leq M_\mathcal{Q}$, the longitudinally polarized partonic cross section decreases faster with increasing $m_c$ than the transversely polarized partonic cross section in the helicity frame. Thus increasing the charm mass results in more transverse polarization. When $p_T > M_\mathcal{Q}$, the longitudinally polarized partonic cross section decreases more slowly with increasing $m_c$ than the transversely polarized partonic cross section. Thus, here increasing the charm mass results in more longitudinal polarization. As $p_T \gg \hat{s}$, $\lambda_\vartheta$ becomes insensitive to $m_c$. Thus the uncertainty in $\lambda_\theta$ is narrower.


\subsection{Sensitivity to feed-down ratios}

We have tested the sensitivity of our results to the feed-down ratios used in our calculations \cite{Digal:2001ue}. Since prompt $J/\psi$ production is dominated by direct $J/\psi$, we vary the feed-down ratio by changing the relative contribution of direct $J/\psi$ and decays from excited states. Thus when the direct fraction, $c_{J/\psi}$, increases, all other $c_{\psi}$ decrease and vice versa. Using the base values of $c_\psi$ in Table~\ref{states} and the reported uncertainty, we vary the feed-down ratios as given in Table~\ref{feed_down_sensitivity}.  Since the polarization of prompt $J/\psi$ production does not vary at central rapidity, we study changes in the polarization by varying the feed-down ratios at $y=0$. The $p_T$-integrated polarization parameter for prompt $J/\psi$ production at $\sqrt{s}=7$~TeV at $y =0$ varies by 0.04 from 0.26 in the helicity frame. This variation is similar to that due to the charm quark mass and renormalization scale variations combined.

\subsection{Sensitivity to diagram weights}
We have tested the sensitivity of our results to diagram weights. As shown in Ref.~\cite{Kniehl:2006sk}, the $\hat{s}$-channel diagram dominates color-octet production at high $p_T$. Turning off the contribution from this diagram by setting $\mathcal{A}_{gg,\hat{s}}=0$ in Eq.~(\ref{matrixelement}) makes a significant difference in polarization as well as the uncertainty band in the high $p_T$ limit. At $5$~GeV, turning off the contribution from the $\hat{s}$-channel diagram reduces the cross section by 70\%. The difference is larger at higher $p_T$. Thus the polarization is more sensitive to charm mass and gives a wider uncertainty band. The polarization parameter at $\sqrt{s}=1.96$~TeV in the rapidity region $|y|<0.6$ in the helicity frame in this case is shown in Fig.~\ref{CDF_HX_tu}. The polarization at low $p_T$ is more transverse compared to Fig.~\ref{CDF_HX}. Instead of becoming slightly transverse at high $p_T$, prompt $J/\psi$ production will remain approximately unpolarized with $\lambda_\vartheta = +0.14^{+0.04}_{-0.14}$ in the helicity frame when the $\hat{s}$-channel amplitude is completely turned off.

\section{Conclusions}

We have presented the transverse momentum and rapidity dependence of the charmonium cross section as well as the the polarization of prompt $J/\psi$ production in $p+p$ and $p$+A collisions in the improved color evaporation model in the $k_T$-factorization approach. We compare the $p_T$ dependence to data at both fixed-target energies and collider energies. We also present $\chi_c$ predictions as a function of $p_T$ at $\sqrt{s}=13$~TeV. We find prompt $J/\psi$ production to be unpolarized at moderate $p_T$ and slightly transverse in the high $p_T$ limit in the helicity frame. We do not observe any rapidity dependence in the polarization in the ranges considered. We report the $p_T$-integrated polarization parameter for prompt $J/\psi$ production at $\sqrt{s}=7$~TeV to be $\lambda_\vartheta = 0.26\pm0.02$ at $y=0$ in the helicity frame. We will study the $p_T$ dependence of bottomonium states in this approach in a future publication.

Since our calculation of the matrix elements is leading order in $\alpha_s$, the high $p_T$ cross section varies strongly with the choice of factorization scale due to the limitations on the uPDFs as $x$ increases. We expect improvements at high $p_T$ when we calculate the cross section to $\mathcal{O}(\alpha_s^3)$ in a future publication. 


\section{Acknowledgments}
We thank B.~Kniehl for the initiation of and encouragement throughout this project. This work was performed under the auspices of the U.S. Department of Energy by Lawrence Livermore National Laboratory under Contract No. DE-AC52-07NA27344 and supported by the U.S. Department of Energy, Office of Science, Office of Nuclear Physics (Nuclear Theory) under Contract No. DE-SC-0004014.


\end{document}